\documentclass[manuscript]{aastex}

\shorttitle{Photospheric and chromospheric magnetic fields}
\shortauthors{Petrie \& Patrikeeva}

\begin{document}

%% LaTeX will automatically break titles if they run longer than
%% one line. However, you may use \\ to force a line break if
%% you desire.

\title{A comparative study of magnetic fields in the solar photosphere and chromosphere at equatorial and polar latitudes}

%% Use \author, \affil, and the \and command to format
%% author and affiliation information.
%% Note that \email has replaced the old \authoremail command
%% from AASTeX v4.0. You can use \email to mark an email address
%% anywhere in the paper, not just in the front matter.
%% As in the title, use \\ to force line breaks.

\author{G.J.D. Petrie \& I. Patrikeeva}
\affil{National Solar Observatory, 950 N. Cherry Avenue, Tucson, AZ 85719\\
Northwestern University, Evanston, IL}

%% Notice that each of these authors has alternate affiliations, which
%% are identified by the \altaffilmark after each name.  Specify alternate
%% affiliation information with \altaffiltext, with one command per each
%% affiliation.

%\altaffiltext{1}{} 
%\altaffiltext{2}{}

%%%%%%%%%%%%%%%%%%%%%%%%%%%%%%%%%%%%%%%%%%%%%%%%%%%%%%%%%%%%%%%%%%%%%%%%%%%%%%%%%
\begin{abstract}

Besides their own intrinsic interest, correct interpretation of solar surface magnetic field observations is crucial to our ability to describe the global magnetic structure of the solar atmosphere.  Photospheric magnetograms are often used as lower boundary conditions in models of the corona, but not data from the nearly force-free chromosphere.  NSO's SOLIS VSM produces full-disk line-of-sight (LOS) magnetic flux images deriving from both photospheric and chromospheric layers on a daily basis.  In this paper, we investigate key properties of the magnetic field in these two layers using more than five years of VSM data.  We find from near-equatorial measurements that the east-west inclination angle of most photospheric fields is less than about $12^{\circ}$, while chromospheric fields expand in all directions to a significant degree.  Using a simple stereoscopic inversion we find evidence that photospheric polar fields are also nearly radial but that during 2008 the chromospheric field in the south pole was expanding super-radially.  We obtain a spatially-resolved polar photospheric flux distribution up to $80^{\circ}$ latitude whose strength increases poleward approximately as cosine(colatitude) to the power $9-10$.  This distribution would give a polar field strength of 5-6~G.  We briefly discuss implications for future synoptic map construction and modeling.

\end{abstract}


%%%%%%%%%%%%%%%%%%%%%%%%%%%%%%%%%%%%%%%%%%%%%%%%%%%%%%%%%%%%%%%%%%%%%%%%%%%%%

\section{Introduction}

The magnetic field of the solar corona is the cause of the most spectacular events in the heliosphere. Since the magnetic field in the corona is very difficult to measure (Judge~1998, Lin, Penn \& Tomczyck~2000, Tomczyk et al.~2008), global atmospheric magnetic field models (Linker et al.~1999, Riley, Linker \& Miki\'{c}~2001, Roussev et al.~2003) and solar wind predictions (Arge \& Pizzo~2000) are usually based on extrapolations from the synoptic measurements of the photospheric magnetic field. Synoptic photospheric field measurements have been routinely available since the 1960s (Howard~1967) and at the NSO since the 1970s (Harvey et al.~1980).  The SOLIS (Synoptic Optical Long-term Investigations of the Sun) Vector Spectromagnetograph (VSM; Keller et al.~2003) has been producing daily full-disk observations of the solar photospheric and the chromospheric magnetic field since August 2003.

%In most of the corona the plasma $\beta <<1$, where $\beta =8\pi p/B^2$ is the ratio of gas and magnetic pressures, $p$ is the plasma pressure and $B=|{\bf B}|$ is the magnetic field strength. This measures the relative influence of the plasma in the dynamics relative to the magnetic field, and so the coronal field is generally approximately force-free.  Furthermore, force-free fields satisfy ${\bf\nabla}\times{\bf B} =\alpha{\bf B}$ where the inverse length scale $\alpha$ is a function of space.  Therefore the influence of currents on the coronal magnetic field is generally small for large-scale structures, and the global corona is adequately modeled using a potential (current-free) solution.  The current-free approximation is well known to be unreliable within active regions and filaments, where strong, organized currents are possible, and at the outer boundary at a few solar radii, where the inertia of the solar wind can drag the weak field far from a force-free state, but the potential model generally works well on global scales.  In practice, PFSS models and MHD models generally produce similar results (Riley et al.~2006).

In most of the corona the ratio of the plasma and magnetic pressures $\beta <<1$.  This parameter measures the relative influence of the plasma in the dynamics relative to the magnetic field, and so the coronal field is generally approximately force-free.  Furthermore, the influence of currents on the coronal magnetic field is inversely proportional to length scale and so is generally small for large-scale structures, and the global corona is adequately modeled using a potential (current-free) solution.  In practice, potential-field source-surface (PFSS) models and magnetohydrodynamics (MHD) models of the global corona generally produce similar results (Neugebauer et al.~1998, Riley et al.~2006).  To extrapolate a potential field solution from boundary data, a single component of the boundary magnetic field is enough to determine the solution fully.  Global MHD models of the solar atmosphere also apply only scalar boundary conditions simply because vector synoptic magnetograms are not yet routinely available.  Such vector magnetograms will be derived from SOLIS VSM images in the future, and from NASA's HMI (Helioseismic and Magnetic Imager) on board SDO (Solar Dynamics Observatory).  PFSS models based on scalar data will continue to be relevant because of the mathematical well-posedness of the potential field problem, the simplicity and ease of use of potential models, and because of their ultimate similarity to much more expensive MHD models.  The general dependence of global atmospheric modeling on scalar boundary data makes the interpretation of these line-of-sight data especially important.  

The PFSS method developed by Altschuler \& Newkirk~(1969) and Schatten et al.~(1969) directly applies as boundary data the magnetic field component along either the radial direction or the observer's line of sight.  Until the early 1990s the LOS component was routinely used in practice.  This approach implicitly relies on the model description being valid throughout the atmosphere, including the layer where the measurements derive from, whereas conditions in the photosphere and corona differ very greatly as far as forces and currents are concerned.  This version of the PFSS model represents well the sector structure of the interplanetary magnetic field and solar wind structure (Wang \& Sheeley~2003).  Furthermore, there is reasonably good agreement between photospheric footprints of open field in the model and observed coronal holes.  A widely-reported deficiency of models based on this approach is that they suffer from weak polar fields: because of projection effects, a nearly radial field close to the pole is barely represented in line-of-sight boundary conditions and is not well reproduced in the model.  This has several important consequences.  The main neutral line of a PFSS solution, identified with the equatorial streamer belt, corresponds poorly with the base of the heliospheric current sheet as reproduced from white-light coronal observations.  These calculated and observed neutral lines correspond well with each other only where they cross the equator.  Elsewhere, the modeled neutral line travels significantly further from the equator, indicating that the poles are too weak, as was first suggested by Pneumann \& Kopp~(1971).  There is also an inconsistency between the observed inclination of polar plumes and the modeled polar field inclination: the plumes are observed to point more equator-ward, indicating that the modeled polar fields are less concentrated than they should be.

Svalgaard et al.~(1978) suggested imposing a polar field distribution whose radial component varies with colatitude $\theta$ as $\cos^8 (\theta )$.  This strengthens the polar fields and moves the neutral line closer to the equator.  Later, Wang \& Sheeley~(1992) proposed a bolder solution to the problem, one that has become standard practice in both PFSS and MHD modeling of the global corona.  Based on a key conclusion of Svalgaard et al.~(1978), that the photospheric magnetic field is approximately radial in general, they advocated abandoning the direct application of line-of-sight measurements as boundary data and instead converting to data for the radial field component by dividing through by the cosine of the heliocentric angle $\rho$, the angle between the radial vector and the line-of-sight direction.  As well as being based on observational evidence, this approach takes into consideration in a simple way the differing physical conditions in the photosphere and corona: the photospheric measurements derive from non-force-free, current-carrying fields oriented in a near-radial direction and separated from the approximately force-free, current-free corona by a thin current layer.  This thin current layer is a mathematical idealization loosely identified with the transition region which is not described in detail by the model.  The validity of applying photospheric radial field information to the radial component of the coronal model relies on the thinness of the transition layer and on the continuity of the field across that layer.  This approach naturally produces an enhanced polar field and, in practice, the calculated neutral lines are in much improved agreement with white-light coronal observations.  Synoptic magnetograms for the radial field component are now constructed by many observatories from line-of-sight measurements of the magnetic field using the assumption that the photospheric magnetic field is approximately radial.  Central to the credibility of this data processing and modeling effort is the observed near-radial orientation of photospheric fields.

In the past, the inclination of photospheric fields has been diagnosed in several ways. Using large time-series of full-disk images from Stanford's Wilcox Solar Observatory (WSO), Svalgaard et al.~(1978) calculated the average line-of-sight field strength across the disk within an equatorial band and found that it varied as $\cos (\rho )$, the cosine of the heliocentric angle, a result consistent with a radial photospheric field.  Using time series from the most poleward apertures of their instrument, they were also able to infer the polar field distribution from the annual modulation of the signal, as mentioned earlier.  Howard~(1974,~1991) inferred the longitudinal east/west component of the photospheric field by comparing the magnetic flux of a region when it was located at equal distances east and west of the central meridian, and found that azimuthal (east/west) tilts of photospheric active-region fields were generally less than $10^{\circ}$ for fields of at least 10~G.  Via a least-squares fit of a simple projection model to Wilcox data, Schrauner \& Scherrer~(1994) found typical tilt angles to be no larger.  Assuming that the center-to-limb variations of continuum contrasts of facular areas is proportional to the angle between the photospheric field and the line of sight, Topka et al.~(1992) drew a similar conclusion.  These findings are in line with physical expectations.  The photospheric magnetic field is dynamically dominated by dense plasma and is confined by the ram pressure of supergranular flows to network boundaries.  The resulting intense flux tubes are much more buoyant than their surroundings and their near-vertical orientation is believed to be due to this buoyancy (Parker~1955,~1966).  Magnetic fields measured using spectral lines formed higher in the atmosphere have been found to behave differently: sunspots and network elements appearing unipolar near disk center acquire as they move limbward a "false" opposite polarity on their limbward side.  This is a signature of a highly inclined, expanding field structure forming canopies at chromospheric heights (Jones~1985).  In contrast to the photosphere, we expect the magnetic field in the chromosphere to be nearly force-free, and therefore not radial. 

Despite its success, the photospheric radial field assumption is often criticized because it makes this strong assumption that the photospheric field is radial, while measurements often show the field to be non-radial, particularly in strong active regions (e.g. Bernasconi~1997).  The picture for non-active fields has also become complicated recently.  Rudenko (2004) pointed out that the radial approximation of the magnetic field in the photosphere is neither confirmed nor contradicted by Svalgaard et al.'s~(1978)  $\cos (\rho )$ distribution of the LOS field described above.  Harvey et al. (2007) discovered the existence of a dynamic horizontal component of the photospheric magnetic field of the quiet Sun. It was known for many years that the quiet Sun contains magnetic fields (Livingston \& Harvey~1971) but this field was not well characterized until recently.  Harvey et al.~(2007) found that the temporal root-mean-square of the dynamic component of the field increases towards the limb, suggesting that these dynamic features are generally nearly horizontally oriented.  This in turn implies that the radial assumption of the line-of-sight magnetic field in the quiet internetwork regions near the limb is wrong. Using Hinode Solar Optical Telescope Spectro-Polarimeter data, Lites et al.~(2008) found ubiquitous dynamic horizontal fields in quiet internetwork regions not coincident with near-vertical, less dynamic fields associated with intergranular lanes.  Here we will investigate the typical orientation of the static components of low-latitude photospheric fields as well as of chromospheric fields, the latter being examined here for the first time.

Besides uncertainty in the direction of surface fields, a further issue is the sensitivity of the models to the fields that are most poorly observed of all: the polar fields.  The two magnetic polar caps are large-scale flux distributions each dominated by a single polarity (Babcock \& Livingston~1958, Babcock~1959, Howard~1972, Timothy et al.~1975, Harvey et al.~1982, Varsik et al.~1999).  While it it known that the polar regions are the dominant sources of fast solar wind during activity minimum (Feldman et al.~1996), the LOS field derived from photospheric line polarization is not strong enough to allow accurate information on the flux distribution near the poles (Varsik et al.~1999, 2002).  Furthermore, the transverse field measurements are not sufficiently sensitive.  In the basic potential field description (Altschuler \& Newkirk~1969, Schatten et al.~1969, Hoeksema~1984, Wang \& Sheeley~1992), on which the more sophisticated nonlinear force-free and MHD models are based, the global large-scale field structure is dominated by the polar dipole component.  The polar dipole is sensitive to the boundary conditions for the polar fields.  Because these polar fields are poorly observed, synoptic magnetogram construction inevitably involves some substitute for unobserved or poorly observed polar fields, such as interpolating estimates from observations at high latitudes (e.g., Liu et al.~2007) or by applying flux transport models that move observed flux into polar regions (e.g., Worden \& Harvey~2000).  Raouafi et al.~(2007) found the latitudinal distribution of flux elements, defined as regions of field with strength greater than 5~G and of size at least a few arcseconds in each direction, to be more concentrated at lower rather than higher latitudes within the polar cap.  On the other hand, the polar flux is generally believed to become stronger as one moves poleward (Svalgaard et al.~1978, Wang \& Sheeley~1988).  In this paper we investigate the tilt and intensity distribution of photospheric and chromospheric polar fields using the SOLIS VSM full-disk images.

The paper is organized as follows.  After introducing the SOLIS VSM images in Section~\ref{solisvsmimages}, we test in Section~\ref{tilttests} the validity of the radial approximation in the photosphere using five years of SOLIS data.  Here we also investigate the typical inclination of the solar chromospheric field.  In Section~\ref{polarfields} we characterize the polar flux in the photosphere and chromosphere.  We determine the typical field inclination close to the poles in both atmospheric layers and give our best estimate of the photospheric polar flux distribution during 2003/8-2008/12.
 
\clearpage
 
\section{SOLIS VSM photospheric and chromospheric line-of-sight magnetic images}
\label{solisvsmimages}

\begin{figure*}[ht]
\begin{center}
\resizebox{0.40\hsize}{!}{\includegraphics*{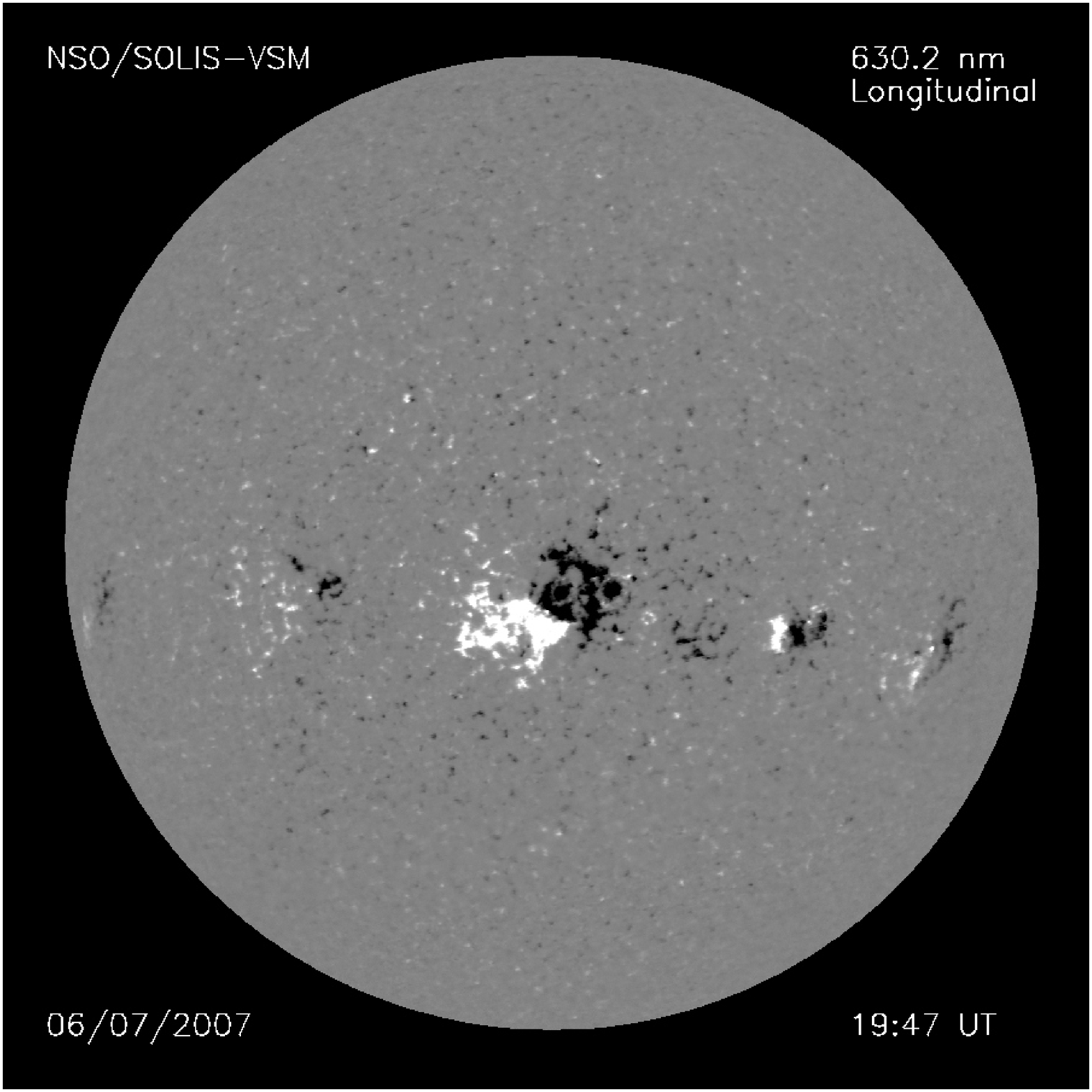}}
\resizebox{0.40\hsize}{!}{\includegraphics*{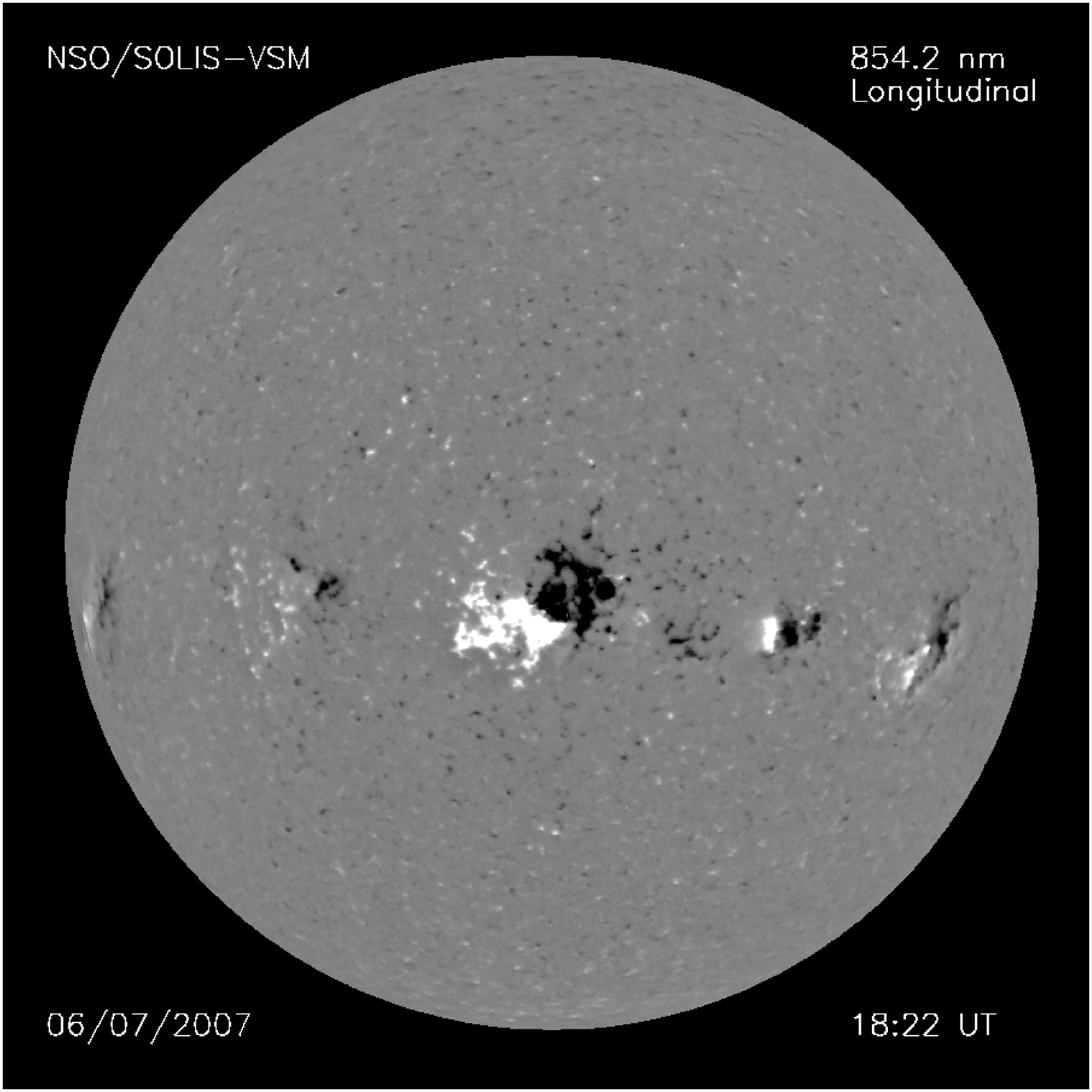}}
\end{center}
\caption{A SOLIS VSM photospheric (6302~\AA) from 6th July 2007 and a chromospheric (8542~\AA) magnetogram from the same instrument on the same day.}
\label{fulldiskimages}
\end{figure*}

Since August 2003, the SOLIS VSM has been producing daily full-disk magnetograms of the solar photosphere using the Fe I (630.2~nm) line and the chromosphere using the Ca II (854.2~nm) line. Two example full-disk images from SOLIS are shown in Figure~\ref{fulldiskimages}: a photospheric example (left) and a chromospheric image from the same day.  White color denotes positive magnetic field (outward from the Sun), and black color denotes negative magnetic field (toward the Sun).  Full-disk observations take about 40 minutes.  The images in Figure~\ref{fulldiskimages} are little more than an hour apart and their similarity is clear.  There are also key differences: magnetic features appear slightly larger and are more diffuse in the chromosphere than in the photosphere.  Furthermore, features close to the limb appear stronger in the chromosphere than in the photosphere.

\begin{figure*}[ht]
\begin{center}
\resizebox{0.89\hsize}{!}{\includegraphics*{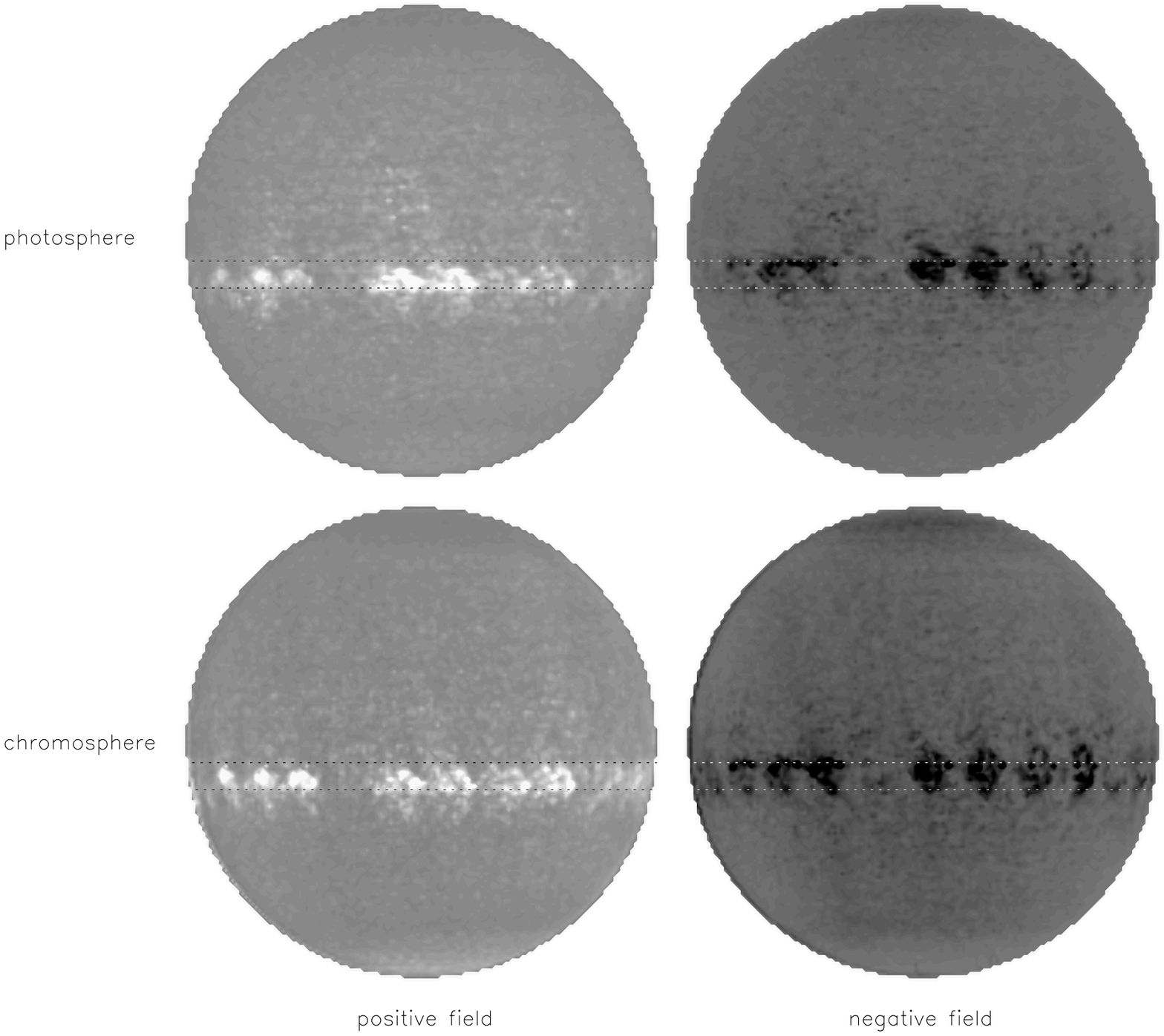}}
\end{center}
\caption{Shown are shaded contour plots of the average magnetic image over a solar rotation (27 days).  White is positive flux and black negative flux, saturated at 10~G.  The top/bottom pictures show photospheric/chromospheric fields.  The left/right pictures show positive/negative fields.  During this rotation the solar tilt angle is close to zero and the active fields are just south of the equator.  Gaps in the active bands in the images correspond to days when no measurements were taken.  Differences between the photospheric and chromospheric fields are described in the text.  The average field strengths across the active bands marked by the dotted lines are plotted in Figure~\ref{avecontourscut}.}
\label{avecontours}
\end{figure*}

\begin{figure*}[ht]
\begin{center}
\resizebox{0.89\hsize}{!}{\includegraphics*{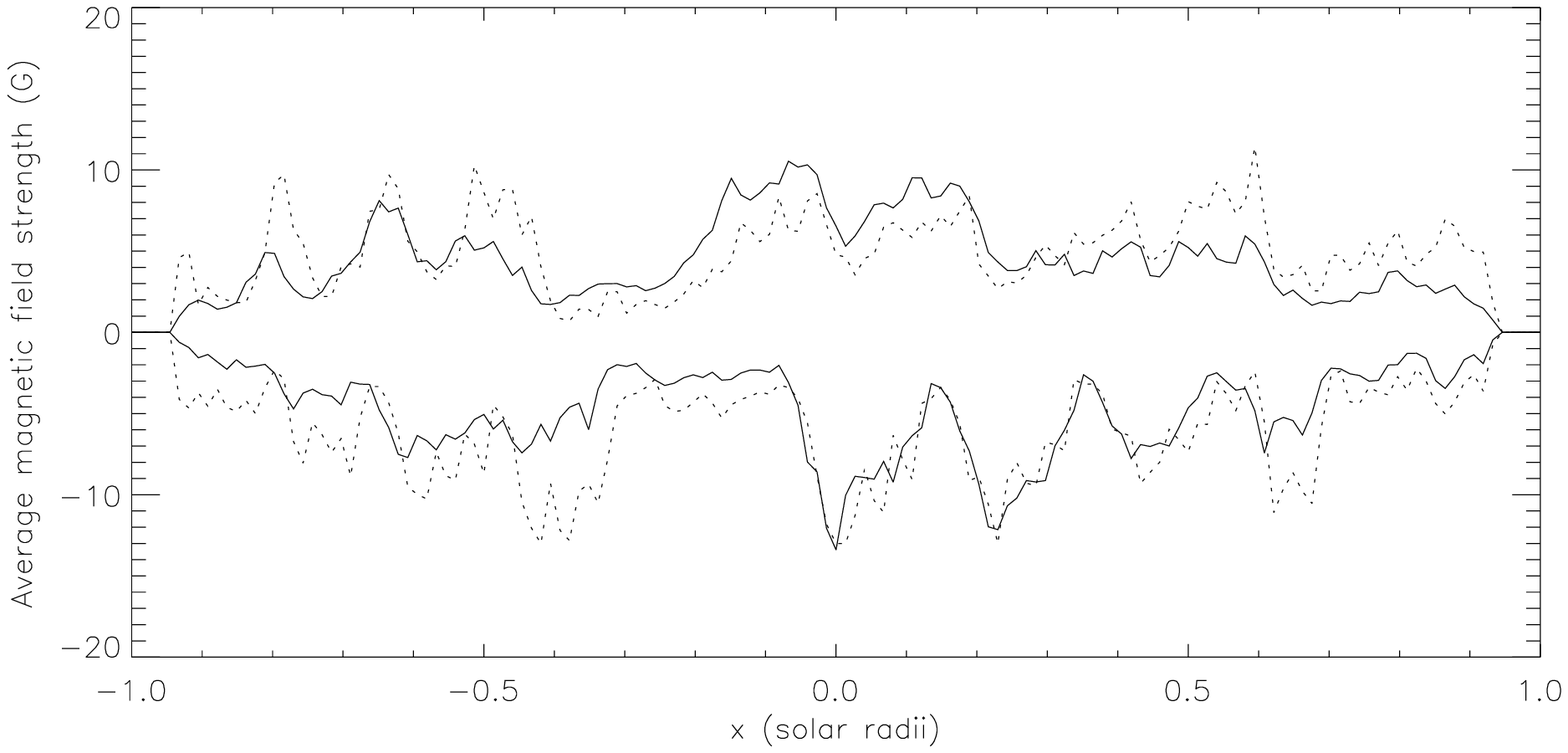}}
\end{center}
\caption{The average field strengths within the active bands of the photospheric (solid lines) and chromospheric (dotted lines) rotation-averaged images in Figure~\ref{avecontours} are plotted here as functions of horizontal position on the disk $x$.  Positive and negative fields are plotted separately.  Thus the solid/dotted graphs correspond to the top/bottom pictures of Figure~\ref{avecontours} and the positive/negative graphs to the left/right pictures of Figure~\ref{avecontours}.}
\label{avecontourscut}
\end{figure*}

We begin with a simple characterization of photospheric and chromospheric fields, based on a straightforward averaging of full-disk line-of-sight images.  We choose a rotation of data (27 days, 5/25/2007 to 6/20/2007) centered at a zero-point of the solar rotation axis's tilt angle relative to us.  Our choice of a time interval late in a solar cycle means that the strong active fields stay close to the equator, simplifying the plots of our averages.  We averaged the total magnetic flux across the full disk of the Sun to find the field strength distribution over the solar disk obtained from the 51 photospheric and 26 chromospheric magnetograms from 5/25/2007 to 6/20/2007.  Separating positive and negative magnetic field in both photospheric and chromospheric averages, we obtained four averages that we show in contour plots in Figure~\ref{avecontours}.
The photospheric fields are not generally stronger than the chromospheric fields. The photospheric field is generally at least as strong near the central meridian than at the limbs, which is not the case for the chromospheric fields, except at the poles. The magnetic field in the photosphere is strongest at the center of the solar disk, which is consistent with a radial photospheric field.  Figure~\ref{avecontourscut} shows this more clearly in plots of field strengths averaged across the active bands marked by dotted lines in Figure~\ref{avecontours}.  These show a marked fall-off from disk-center towards the limbs in the photospheric data whereas the chromospheric plots do not show such a clearly defined trend.  In Figure~\ref{avecontours} the strengths of the photospheric and chromospheric polar fields have maxima near the central meridian, perhaps because the fields in both layers have near-vertical structure and simple topology.  During 2007 the Sun is quiet, and the solar magnetic field resembles a dipole with a simple, well-defined magnetic field strength maximum at the poles.  This is how the field distribution appears in the plots.  They are more pronounced in the chromospheric plots than in the photospheric plots, suggesting that the polar fields expand super-radially.  Away from the poles, we expect the photospheric field to be nearly vertical, but the chromospheric magnetic field to have a more complex structure, which seems to be the case according to these plots.

 \clearpage

\section{Heliocentric angle dependence and derived tilts of photospheric and chromospheric line-of-sight fields}
\label{tilttests}

The simple averaging procedure of the previous section indicates that there are key differences between the photosphere and chromosphere in the field observations.  To investigate these differences more thoroughly, we expand our investigation to all available photospheric and chromospheric observations: once or twice daily since August 21 2003.

\begin{figure*}[ht]
\begin{center}
\resizebox{0.75\hsize}{!}{\includegraphics*{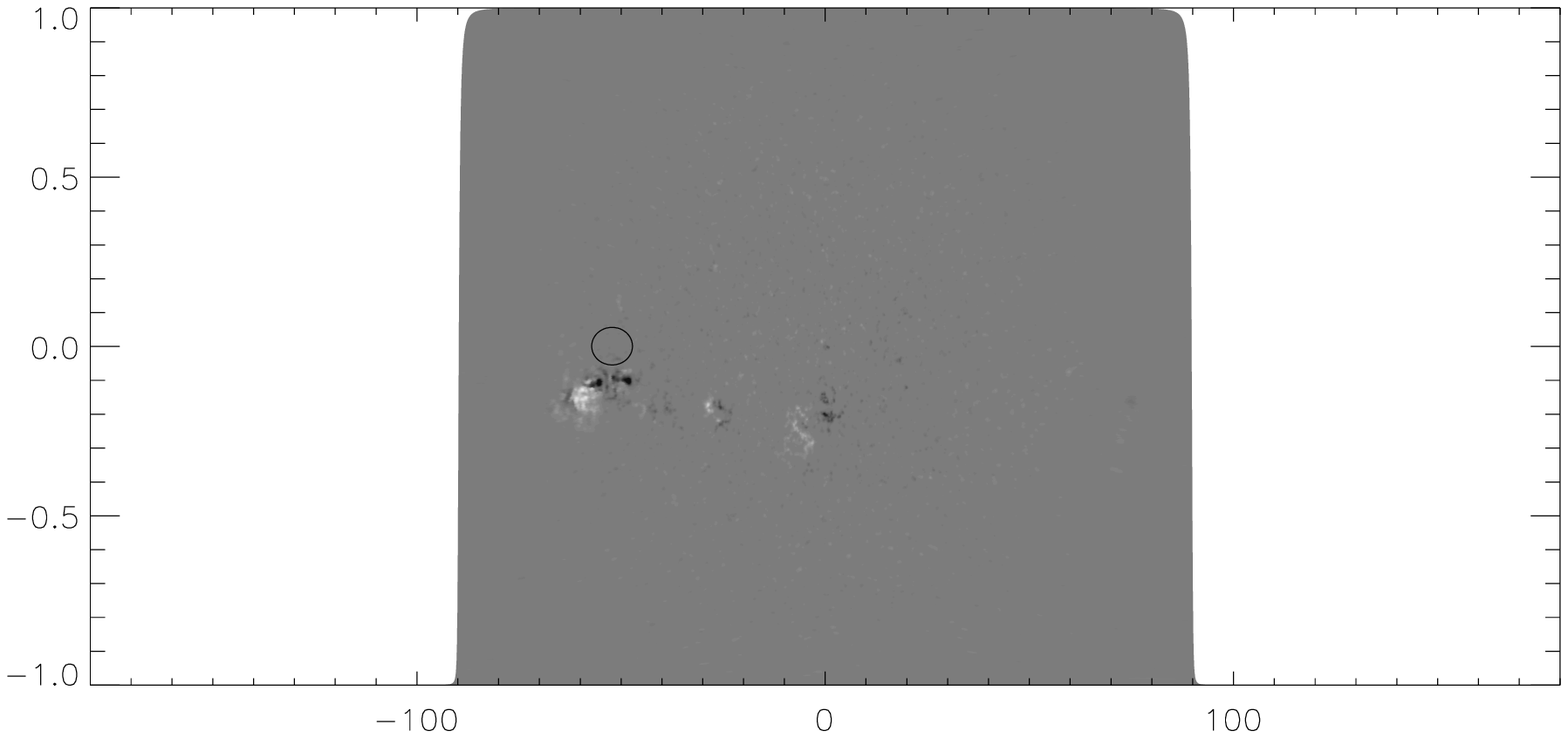}}
\resizebox{0.75\hsize}{!}{\includegraphics*{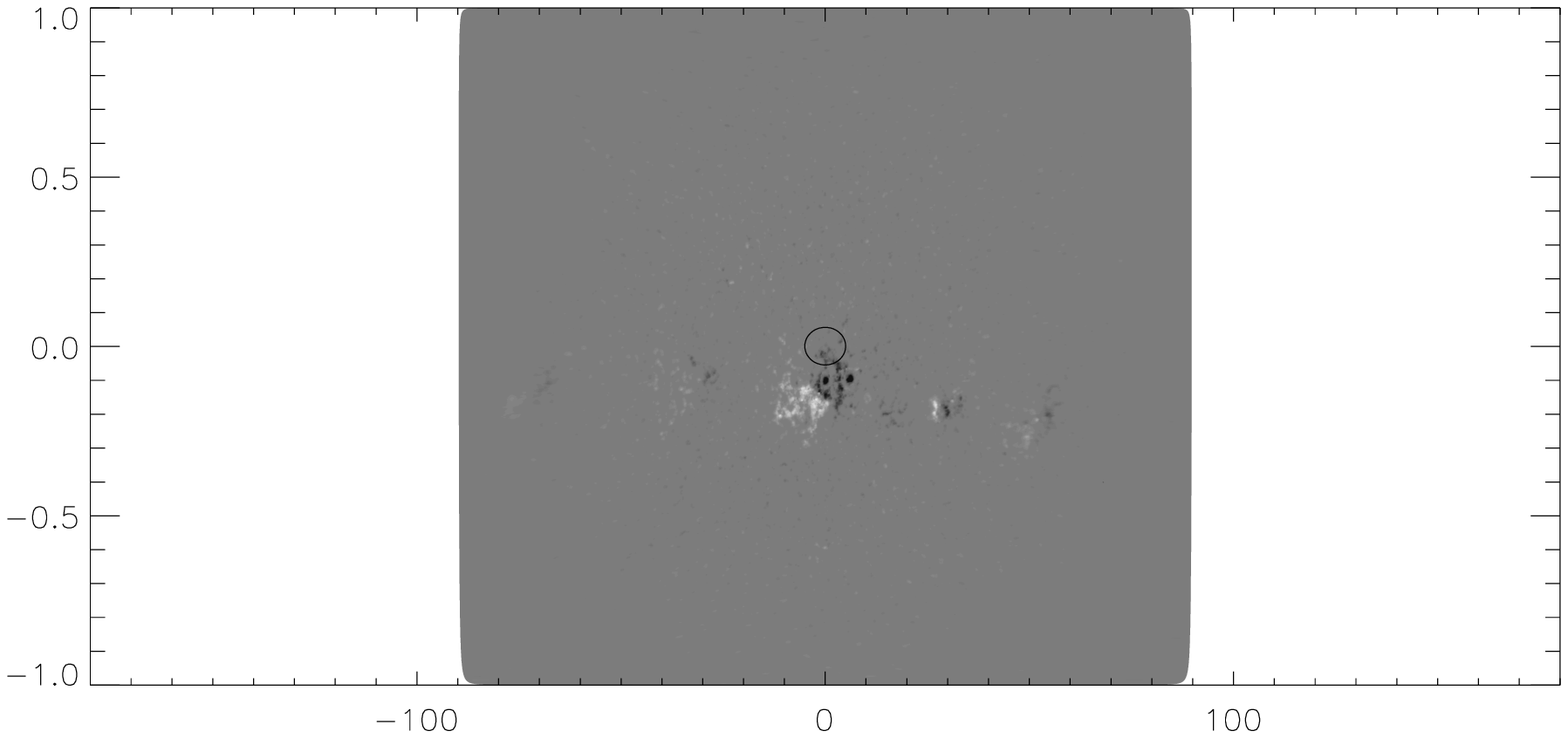}}
\resizebox{0.75\hsize}{!}{\includegraphics*{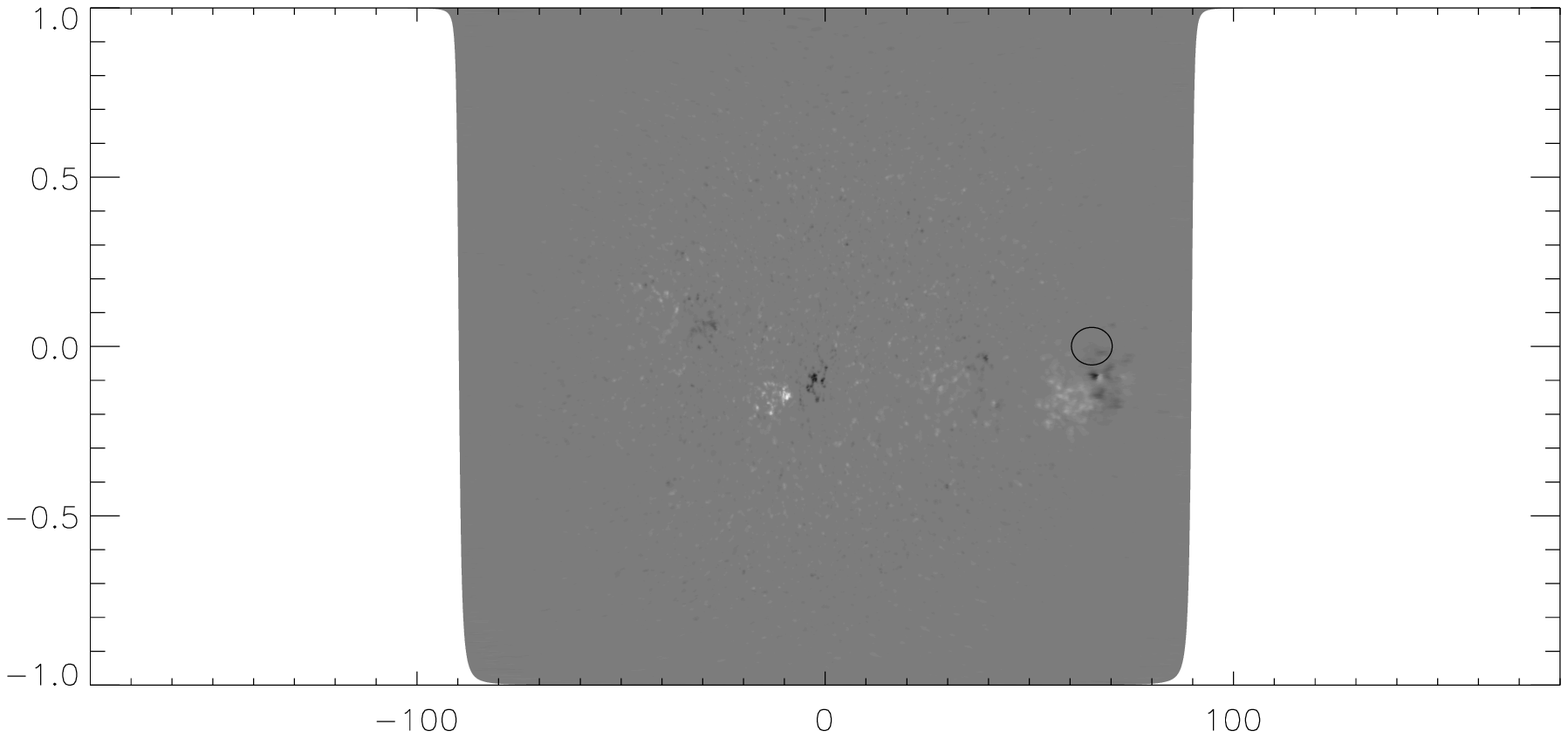}}
\end{center}
\caption{Remapped photospheric magnetograms in heliographic coordinates from 6/3/2007 (top), 6/7/2007) (middle) and 6/12/2007 (bottom).  The middle remap is derived from the photospheric image shown in Figure~\ref{fulldiskimages}.  The circle in the middle picture has diameter $10^{\circ}$ and is centered at the location of disk center, $(0,B_0)$ in heliographic coordinates.  The locations of this circle on 6/3/2007 and 6/12/2007 are shown in the other two pictures.}
\label{remaps}
\end{figure*}

Our method is as follows.  In every magnetogram we identify the region within $10^{\circ}$ of disk center and then trace the evolution of the average line-of-sight flux density of this bundle as it traverses the disk.  To simplify this calculation, we transform every image into heliographic coordinates.  A circle in the heliographic plane does not change shape under rigid solar rotation.  Moreover, complications associated with the tilt angle, which we denote by $B_0$, of the solar rotation axis with respect to the ecliptic plane are eliminated: solar rotation moves strictly in the longitudinal direction in heliographic coordinates.  The Sun rotates differentially about its axis with period approximately 27~days, as observed by the motion of the sunspots and other photospheric features. We  take a circle of $10^{\circ}$ in diameter on the disk center, and identify the region covered in this circle in earlier and later magnetograms assuming a rotation period of 27.27 days, the rotation period adopted in constructing standard synoptic maps. We cannot expect the circle to capture the same flux from image to image.  Apart from the fact that flux is continually emerging and being annihilated in the photosphere, the various flux transport processes will carry some flux across the circle boundary.  However, a circle as large as $10^{\circ}$ in diameter can be expected to contain much the same flux from image to image much of the time.  Figure~\ref{remaps} shows three solar images in heliographic coordinates with the $10^{\circ}$ showing such a circle's location at three different times. On 6/7/2007 (Figure~\ref{remaps}, middle) the $10^{\circ}$ circle is at the center of the solar disk and almost at the equator in the remap. This is because the value of $B_0$ for this day is approximately $0.0934^{\circ}$, which is very small. The other plots in Figure~\ref{remaps} (top and bottom) show the circle on the solar surface about four days before and five days after the central magnetogram as it moves along the latitude of the solar disk center. The circle has equal heliographic area throughout the disk passage.  This makes it simple to calculate the line-of-sight flux through the circle as it traverses the disk.

\clearpage

\clearpage

\begin{figure*}[ht]
\begin{center}
\resizebox{0.49\hsize}{!}{\includegraphics*{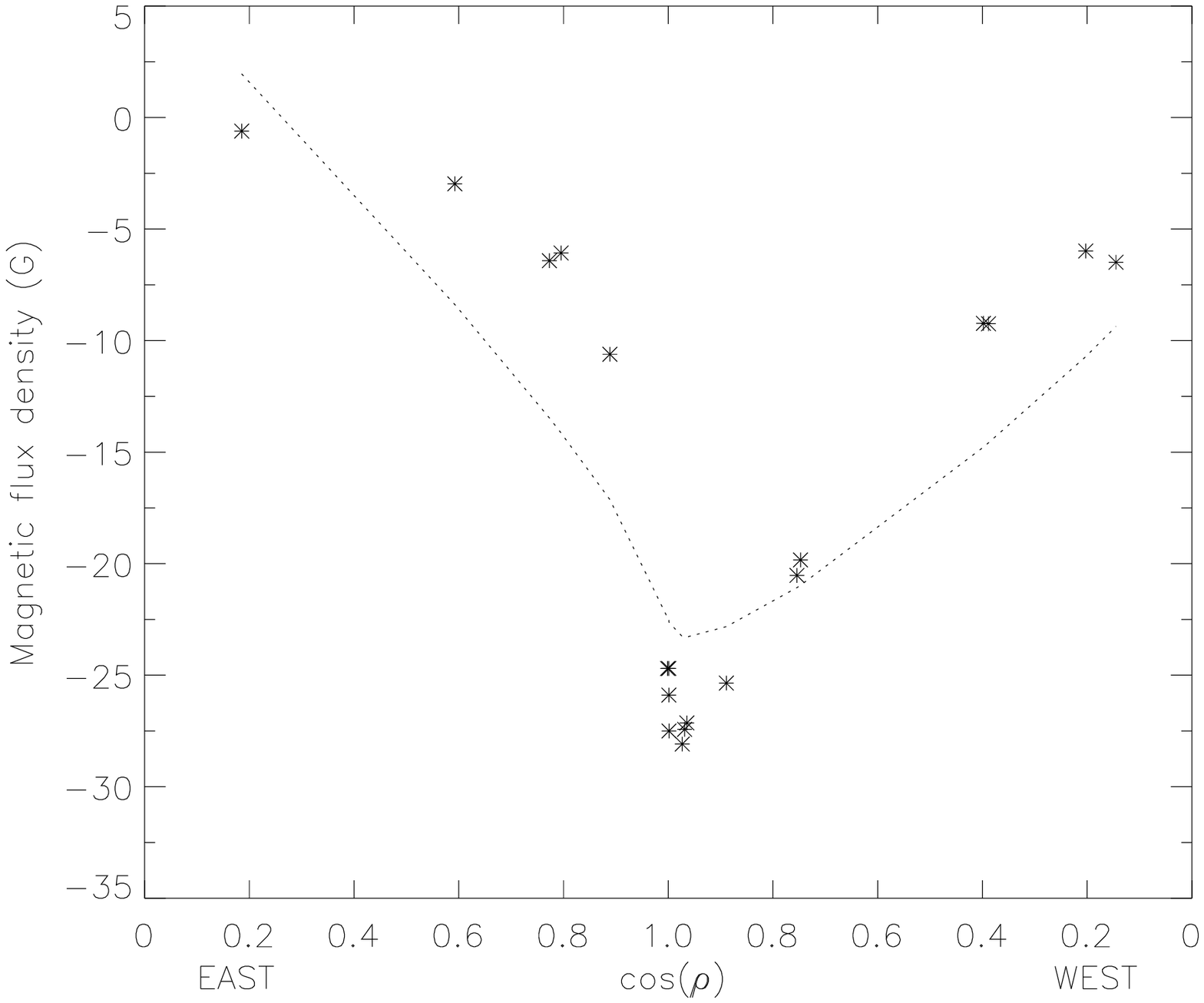}}
\resizebox{0.49\hsize}{!}{\includegraphics*{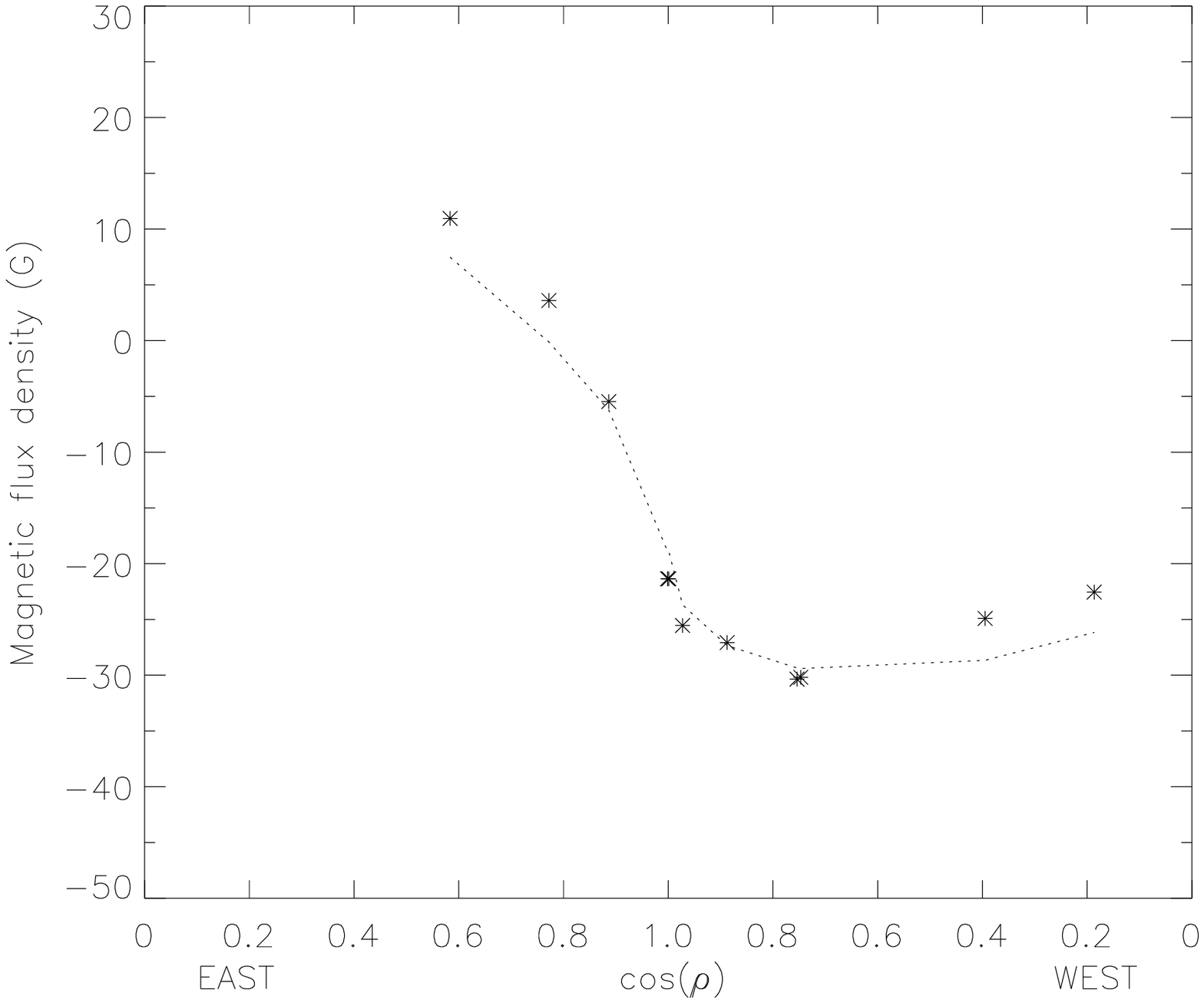}}
\end{center}
\caption{The 'x' symbols show the average magnetic intensity within the $10^{\circ}$ circle of Figure~\ref{remaps} as it traverses the solar disk as a function of the cosine of heliocentric angle $\rho$.  Examples are shown for the photosphere (left) and the chromosphere (right).  These examples apply as their reference magnetograms the images shown in Figure~\ref{fulldiskimages}.  The dotted lines show how the line-of-sight components of the best-fitting field vectors to the measurements as functions of $\cos (\rho )$.  See the text for details.}
\label{symmetryex}
\end{figure*}

\clearpage

\begin{figure*}[ht]
\begin{center}
\resizebox{0.49\hsize}{!}{\includegraphics*{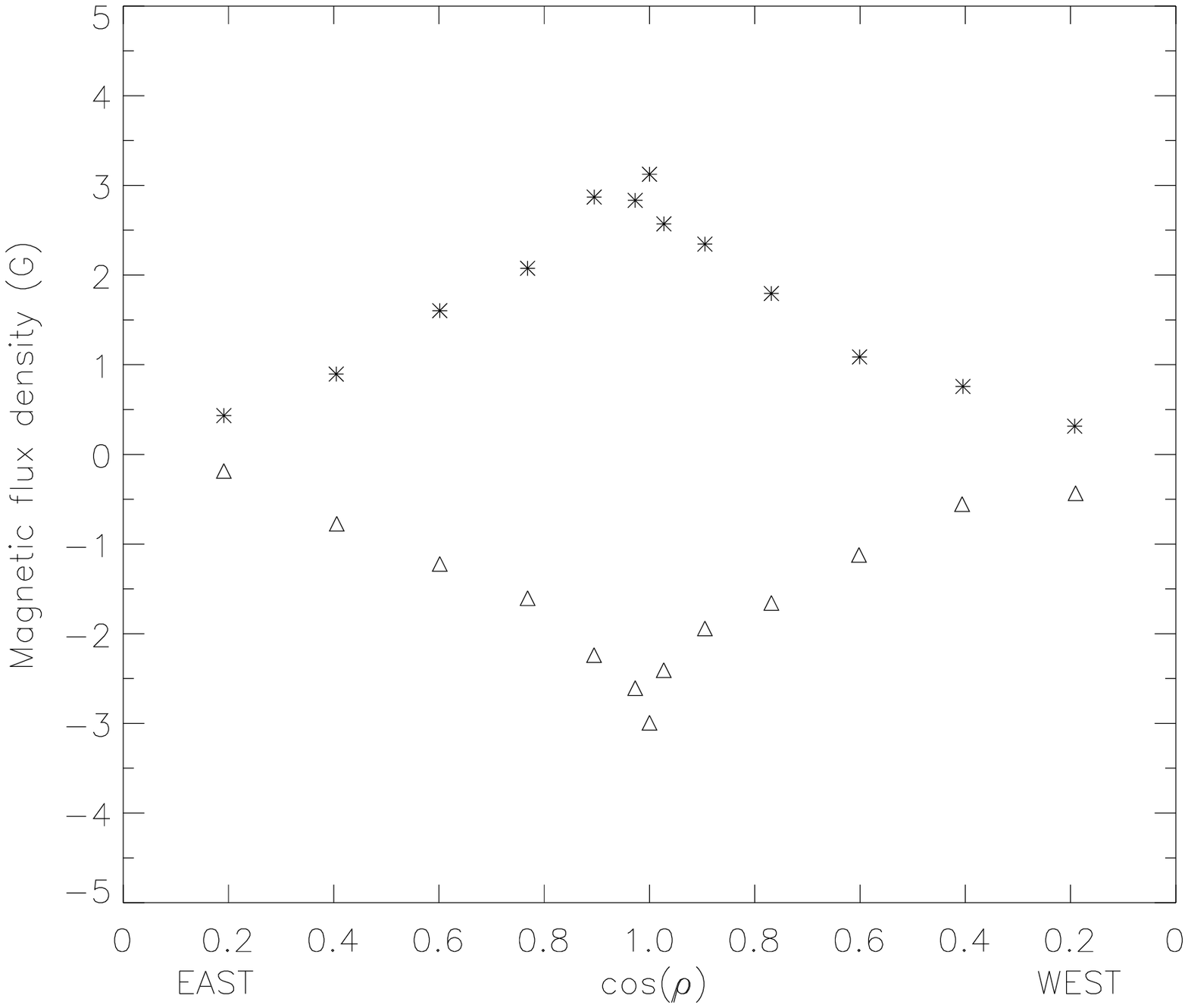}}
\resizebox{0.49\hsize}{!}{\includegraphics*{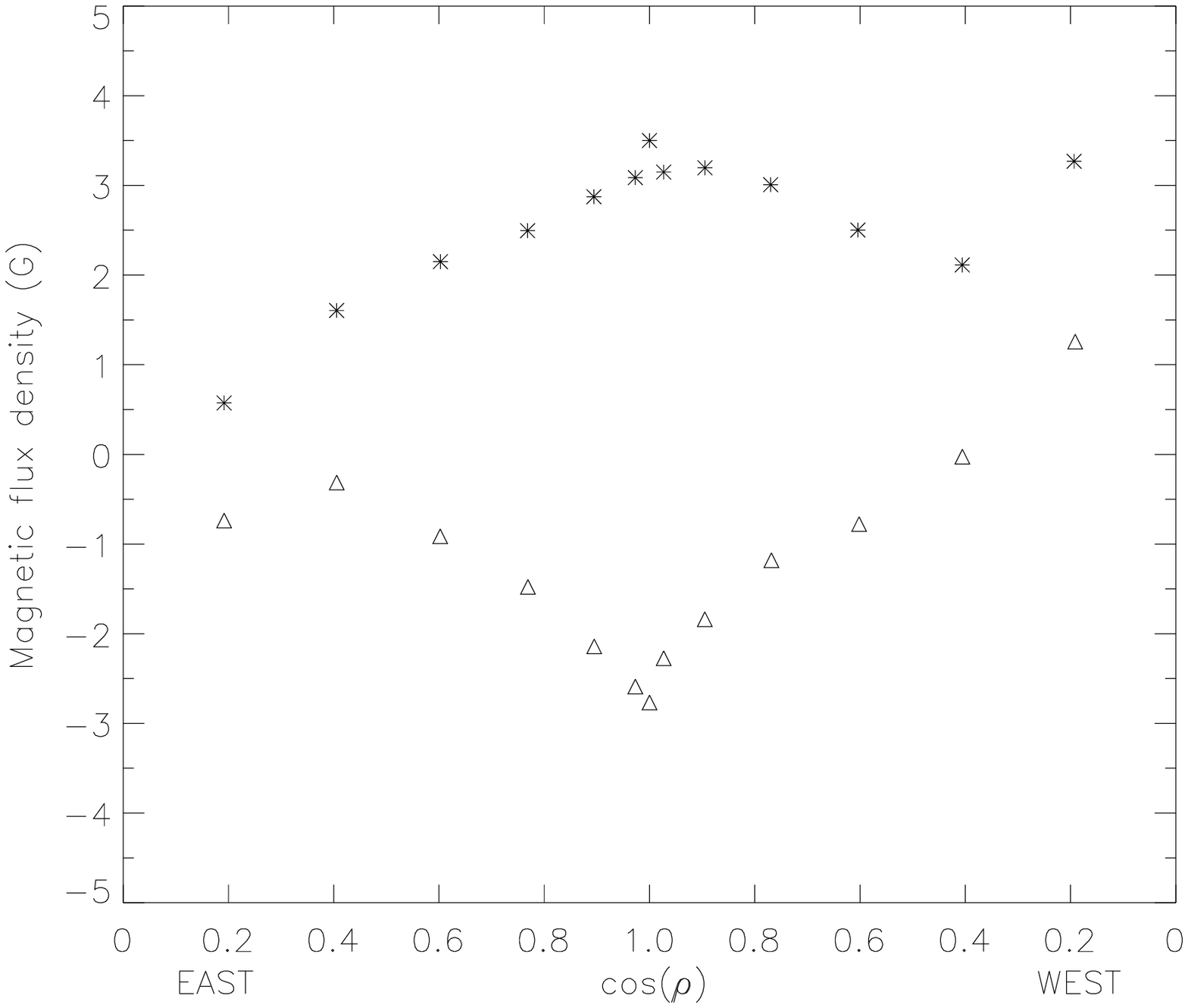}}
\resizebox{0.49\hsize}{!}{\includegraphics*{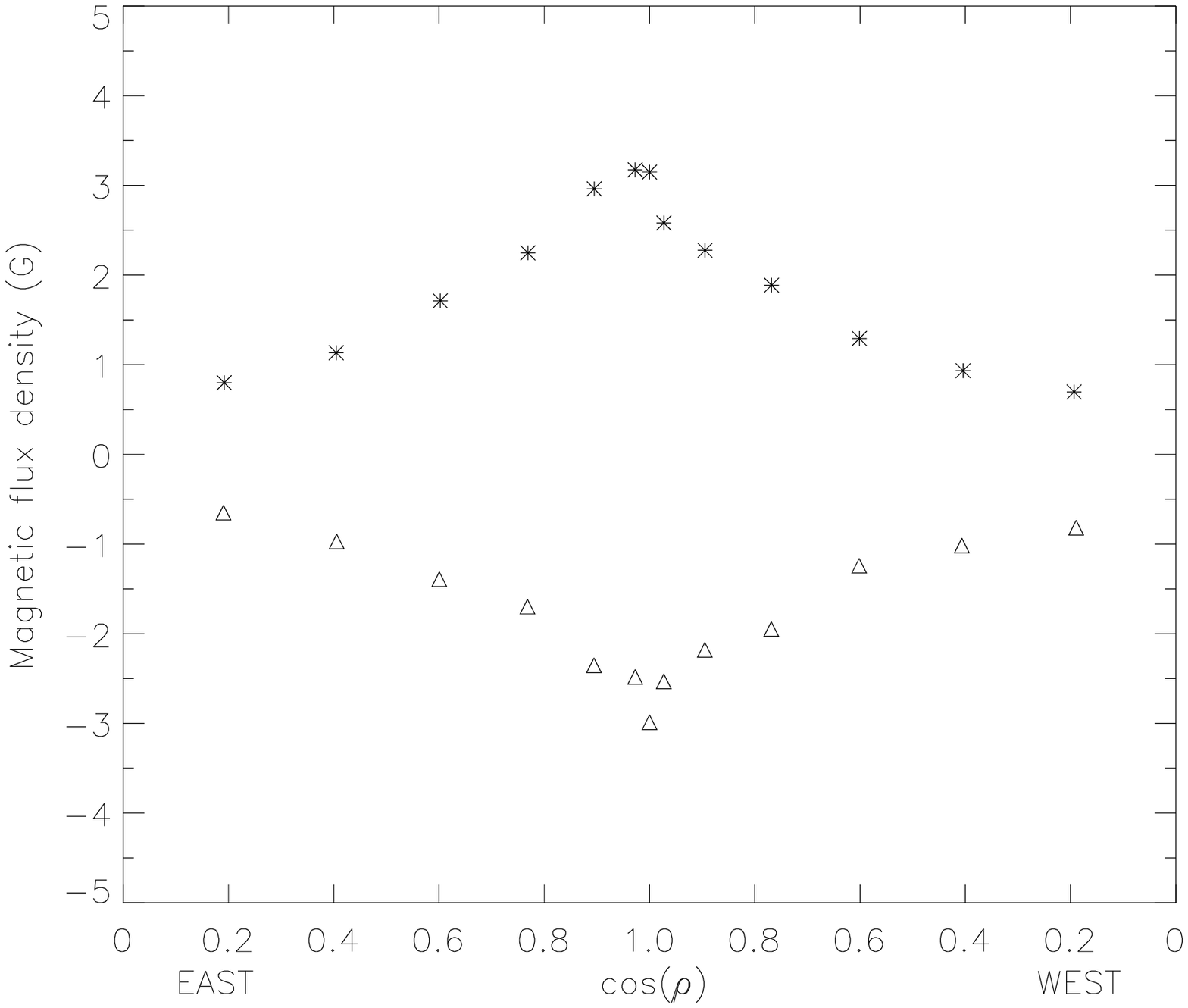}}
\resizebox{0.49\hsize}{!}{\includegraphics*{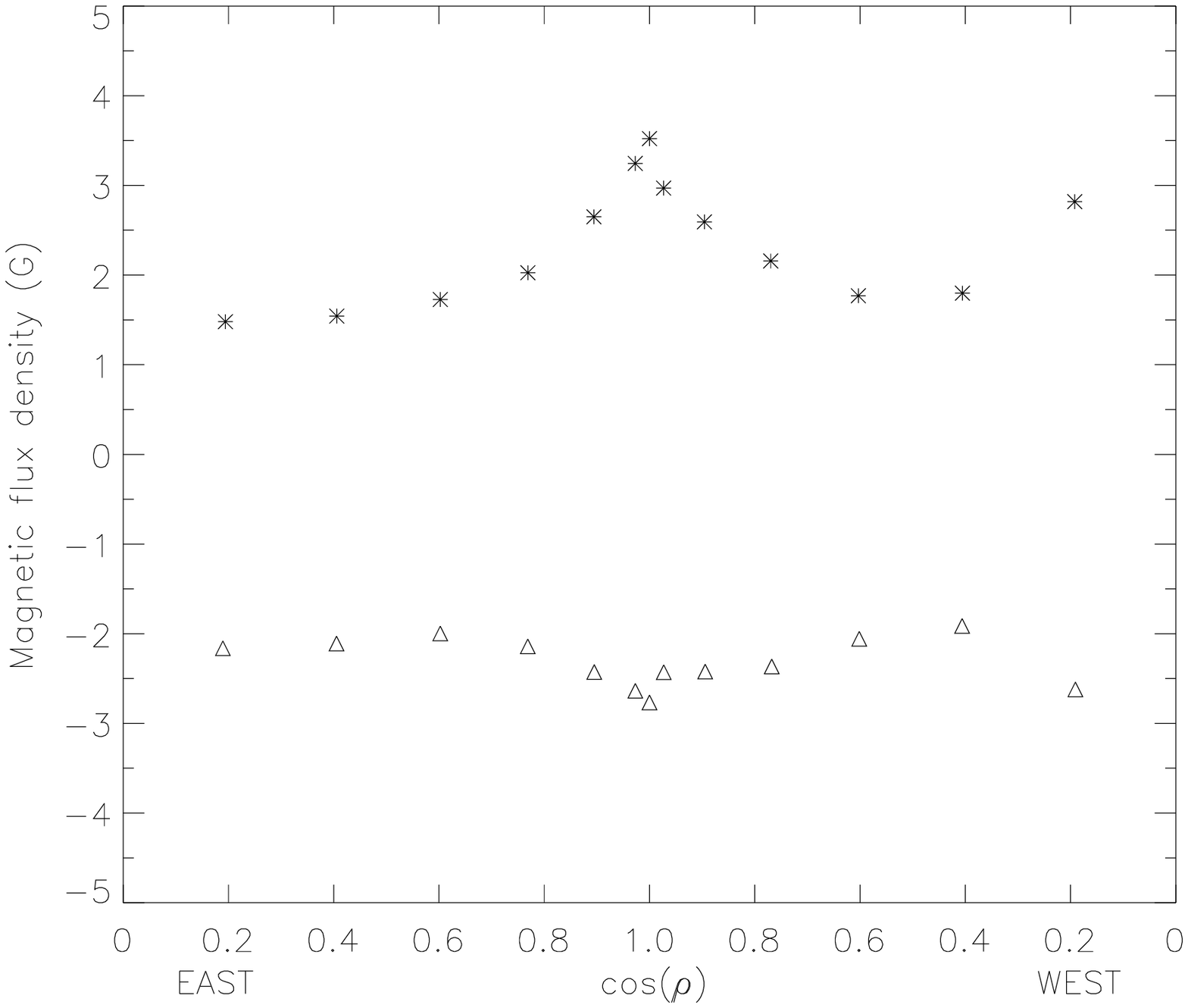}}
\end{center}
\caption{Average magnetic intensity as a function of cosine of heliospheric angle for all examples in the photosphere (left) and the chromosphere (right).  Stars denote averages of positive fields and triangles negative.  The measurements are binned according to cosine(viewing angle) and disk-center net polarity (top pictures) or local polarity (bottom pictures).  The photospheric fields show a clear dependence on heliospheric angle while the chromospheric fields do not.  Because these averages involve a wide range of fields, from weak fields of less than 1~G to a few cases of fields averaging around 100~G, the standard deviations are very large: from about 2~G at the limbs to between 5 and 10~G at disk center.}
\label{aveflux}
\end{figure*}

\clearpage

\begin{figure*}[ht]
\begin{center}
\resizebox{0.49\hsize}{!}{\includegraphics*{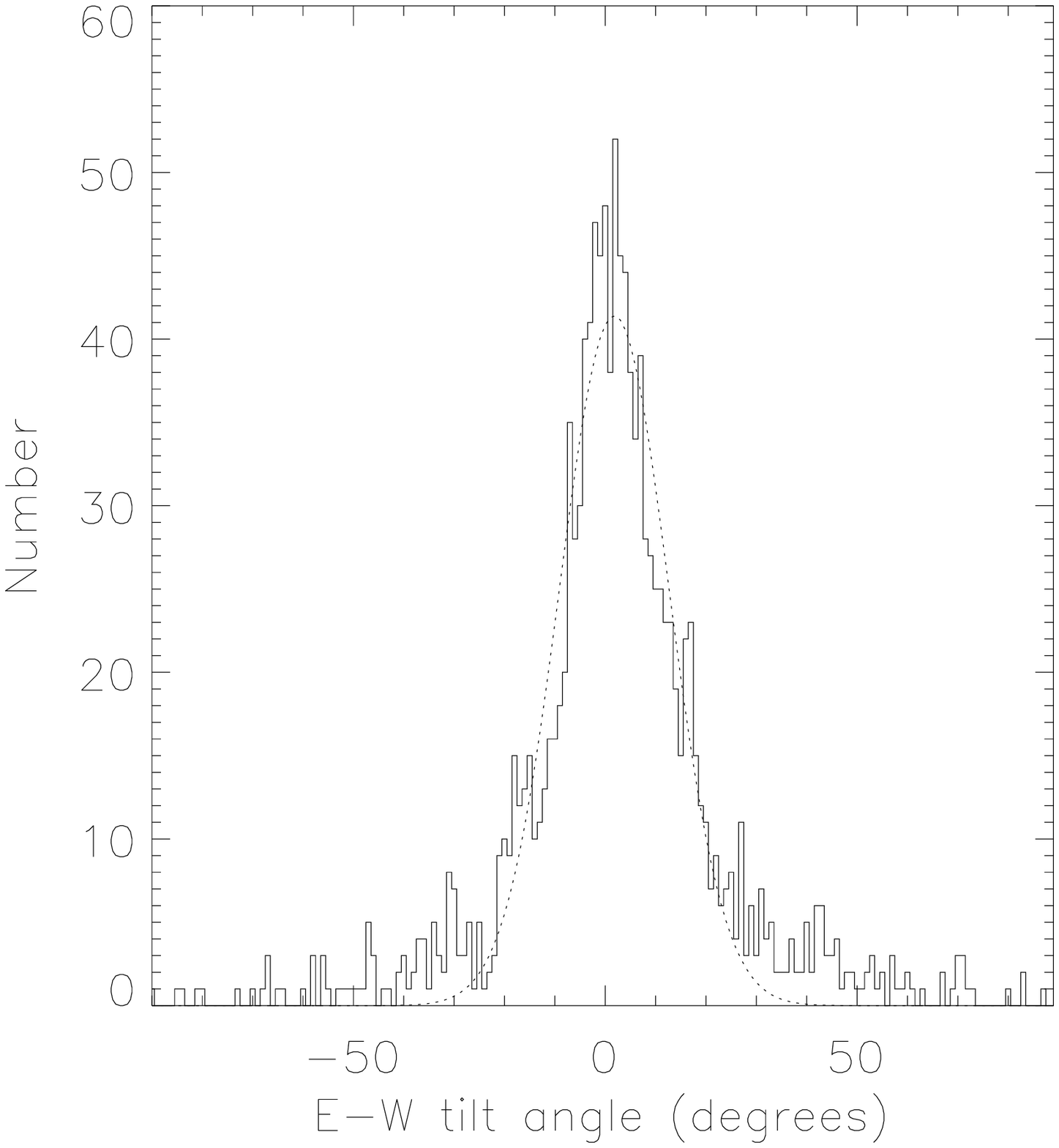}}
\resizebox{0.49\hsize}{!}{\includegraphics*{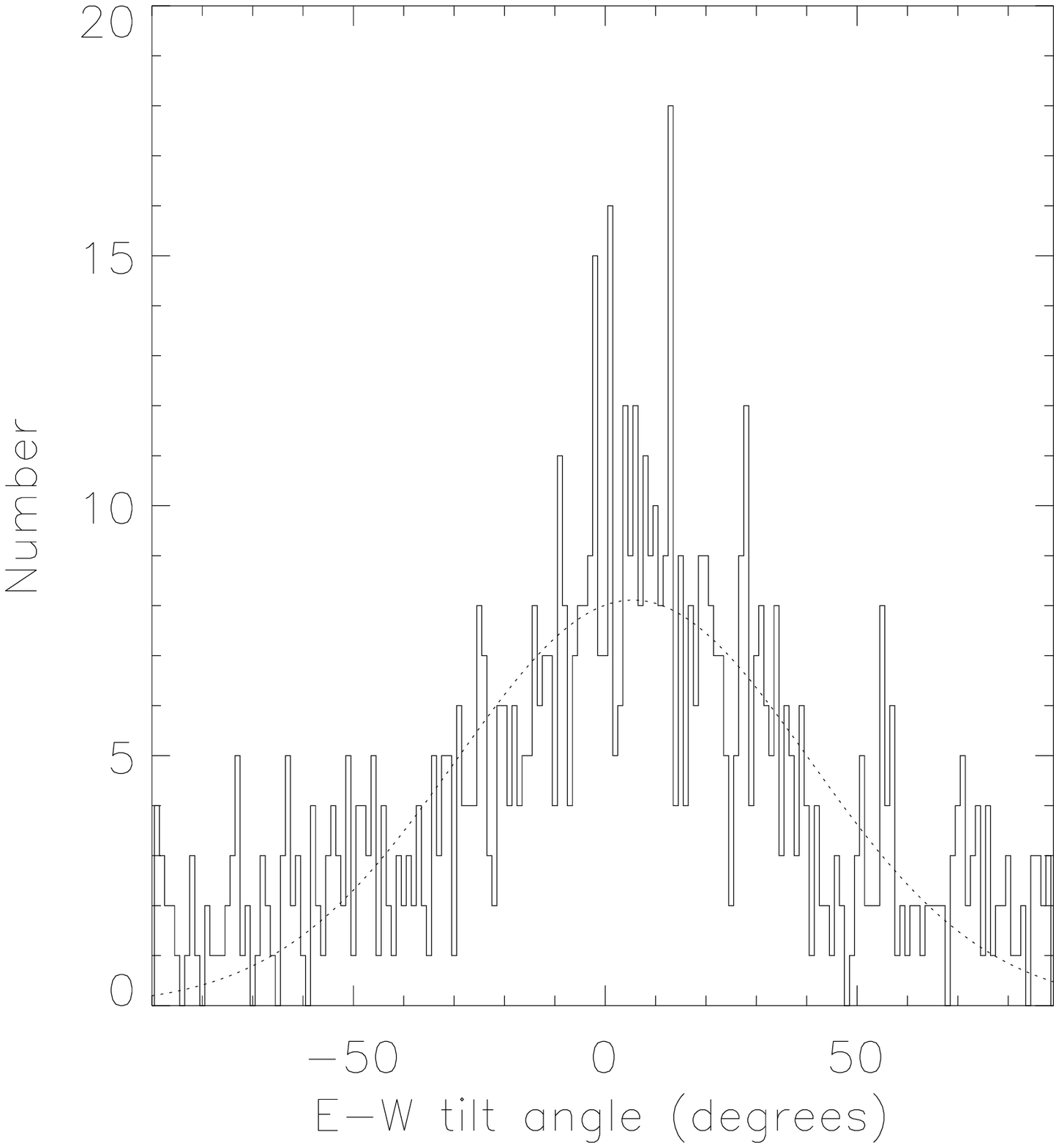}}
\end{center}
\caption{Histograms of photospheric (left) and chromospheric (right) east-west tilt angles and best Gaussian fits.  According to the Gaussian fits, the photospheric fields have tilt angle $1.8\pm 10.8^{\circ}$ and the chromospheric fields $5.5\pm 35^{\circ}$.  These histograms are consistent with a field structure that is nearly radial in the photosphere that expands in a greater variety of directions at chromospheric heights.}
\label{hists}
\end{figure*}

\clearpage

\begin{figure*}[ht]
\begin{center}
\resizebox{0.89\hsize}{!}{\includegraphics*{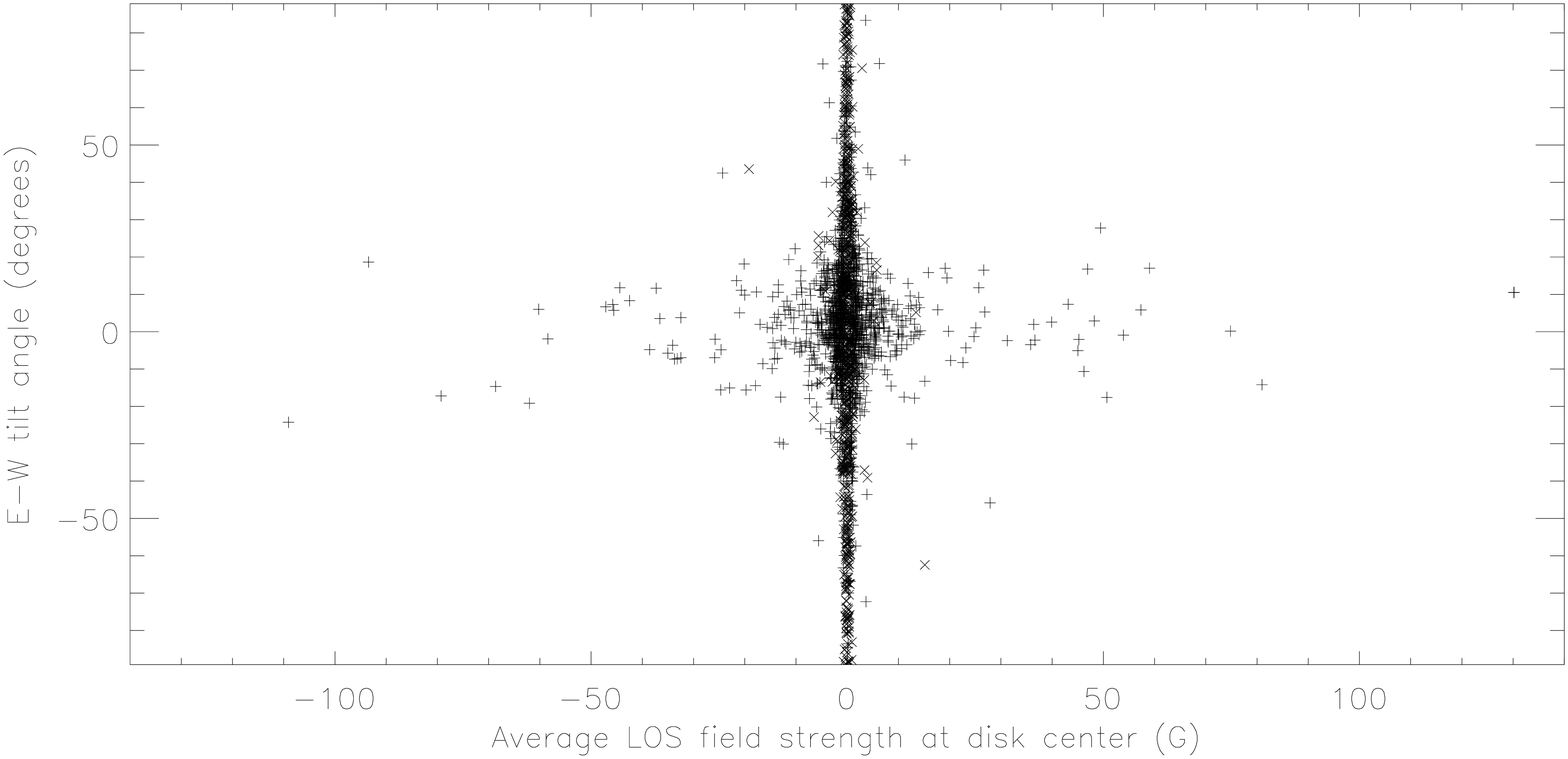}}
\resizebox{0.89\hsize}{!}{\includegraphics*{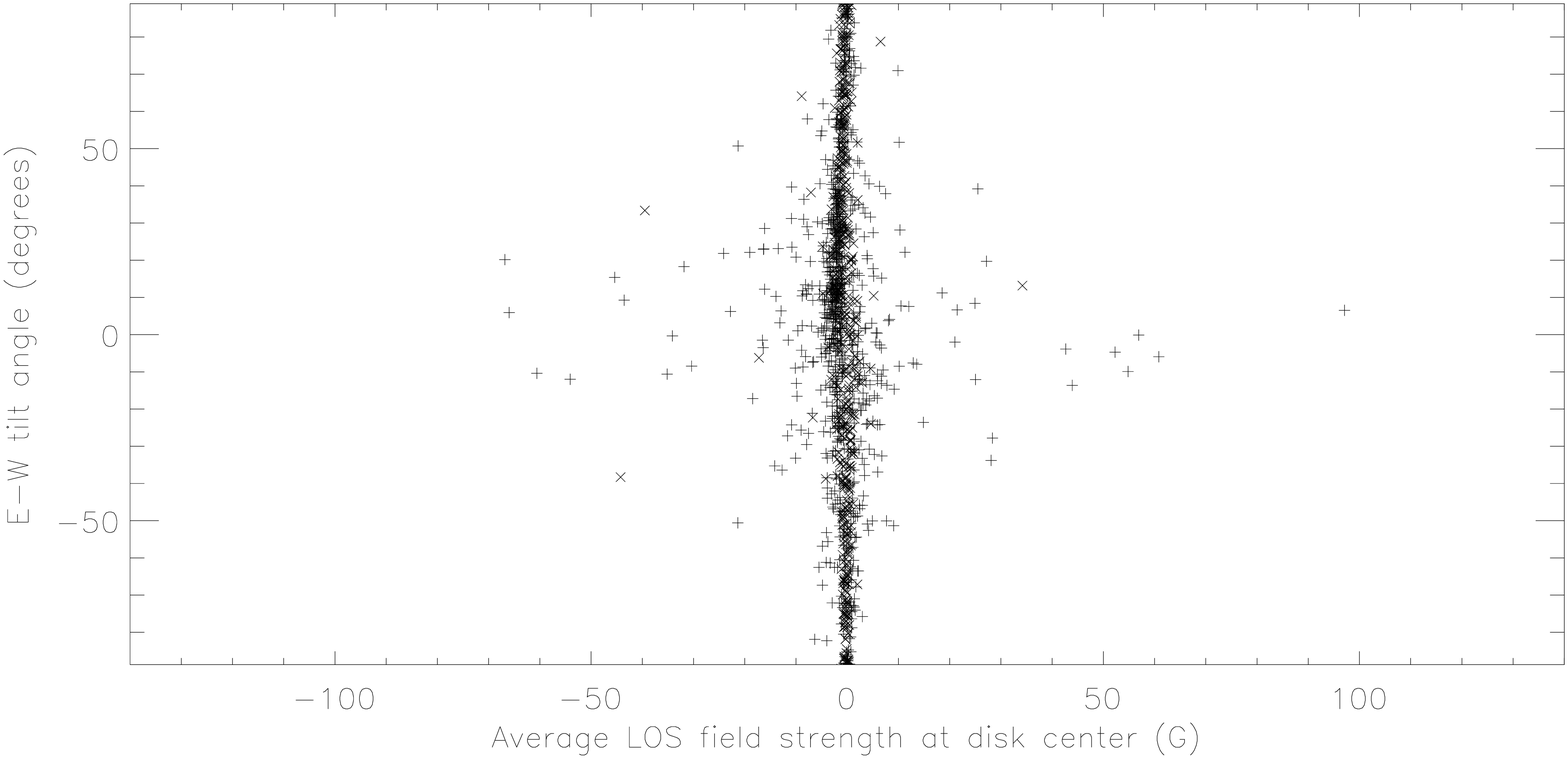}}
\end{center}
\caption{Scatter plots of the disk-center LOS field strength against E-W tilt angle for the photospheric (top) and chromospheric (bottom) examples.  Examples where good-quality curve fits were possible are marked with a + and the remaining examples with three or more data points are marked with a X.  Among the 2008 photospheric cases, satisfactory curve fits were possible for 1323 out of the 2002 cases with three or more data points.  Among the 1268 chromospheric cases, satisfactory curve fits were possible for 747 out of the 1254 cases with three or more data points.}
\label{qualscatter}
\end{figure*}

We plot in Figure~\ref{symmetryex} (left) the average photospheric magnetic flux density inside the $10^{\circ}$ circle as a function of $\cos\rho$, separating measurements east and west of central meridian.  The dependence on $\cos\rho$ is very similar to the dependence on longitude (time) because the $B_0$ angle is so small on 6/7/2007 ($0.0934^{\circ}$).  For the chosen central magnetogram at 6/7/2007 the net magnetic flux inside the circle of $10^{\circ}$ along the equator is negative (there is a feature of negative polarity inside the circles in Figure~\ref{remaps} that appears strongest on 6/7/2007, shown in the middle plot), and it is largest at the center of the solar disk (Figure~\ref{remaps}, middle). The flux density decreases approximately linearly with distance from the center of the disk. This observation is consistent with radial orientation of the magnetic field in the photosphere.  We will discuss the dotted lines shortly.

We calculated the unsigned flux averaged over each 10-degree circle as it rotated across the disk of the Sun.  We did this for every such circle found in the 5-year data set of 2008 photospheric and 1268 chromospheric magnetograms.  We plot the results in Figure~\ref{aveflux} (top pictures), splitting the sample of 10-degree regions into bins, with each bin
containing values from regions having both the same disk-center net polarity and heliocentric
angle $\rho$.  A purely radial field would give four straight lines forming a rhombus.  Based on these plots, the photospheric field seems to be approximately radial and the chromospheric less so.  However, the LOS component of a non-radial field can change its sign on its way across the disk.  Therefore, when the field value at disk center is used as the selection parameter (top pictures), bins for positive/negative fields can contain fields of opposite polarity away from central meridian.  Averaging of mixed-polarity LOS fields at the limbs might involve a significant degree of cancelation and give the false impression of weak LOS fields there.  In this way, non-radial fields whose LOS component changes sign during disk passage can appear to be radial  according to this test, a point noted by Rudenko~(2004).  In the bottom plots we instead split the sample of 10-degree regions into different bins: each of these bins
contains regions having both the same local net polarity and
angle $\rho$.  This binning has the (somewhat unsettling) effect
that the same 10-degree region, as it rotates across the disk, might
appear variously in the positive or negative bins, depending on the net polarity of
the region at the time of the observation.  However, comparison of the top and bottom plots of Figure~\ref{aveflux} does reveal that such LOS field sign changes appear to be significant in the chromosphere but not in the photosphere.  The top and bottom plots of a purely radial field would be identical.  These plots together amount to evidence that the photospheric field is nearly radial in general while the chromospheric field is not.

The presence of scatter in the photospheric and chromospheric plots could be due to the fields' dynamical evolution: fields move in and out of the $10^{\circ}$ circle during its passage from limb to limb, and the dynamics change the field strength and orientation.  Furthermore, the photospheric field is not generally radial in strong active regions (Bernasconi~1997), which must have some influence on the photospheric results. The standard deviations associated with these averages are very large because of the large range of fields being summed together, from weak fields less than 1~G in strength to active kilogauss fields pushing $10^{\circ}$-region averages up to around 100~G.  Interpretation of these standard deviations is not straightforward because these are not independent measurements of a single quantity.  Nevertheless, the average photospheric field orientation seems to be close to radial in Figure~\ref{aveflux}.  Comparing our the photospheric and chromospheric results, the plot of average magnetic flux as a function of the line-of-sight angle in the chromosphere is more scattered, and the magnetic flux  does not show a simple, well-defined dependence on $\cos \rho$ while the photospheric results are consistently close to linear dependence on $\cos (\rho )$.  The photospheric results are consistent with a  radial field structure, whereas the chromospheric field shows no particular preferred orientation and appears to expand in all directions to a significant degree.  As was the case with the fields studied by Svalgaard et al.~(1978), there is asymmetry in the top plots of Figure~\ref{aveflux}, particularly the right plot of chromospheric fields.  This asymmetry suggests the presence of a weak azimuthal field in the west-east direction.

Returning our attention to Figure~\ref{symmetryex} the dotted curves represent the best-fitting magnetic vector $(B_r , B_{\phi})$ through all data points in each case.  It is created by placing radial and azimuthal unit vectors along the trajectory across the disk described by our circle of flux, projecting these vectors along the LOS direction and graphing the projected LOS flux distributions as functions of $\cos (\rho )$.   The derived vector components are the coefficients of the best-fitting linear combination of unit radial and azimuthal LOS curves to the data.  By finding the best-fitting linear combination of radial and azimuthal fields to each of our 3000 or so graphs of the type shown in Figure~\ref{symmetryex}, we obtained an estimate of the east-west tilt angle of the field for each example.  See Schauner \& Scherrer~(1994) for a different approach, also adopted by Ulrich et al.~(2002).  Here we focused on fields that pass through disk-center so that each data set contains a pure radial field measurement.  We cannot invert for the poloidal (north-south) component, $B_{\theta}$, using this technique because its direction is nearly perpendicular to the line of sight near the equator and therefore its LOS projection does not contribute significantly to these nearly equatorial LOS measurements.  Because these calculations exclude poloidal fields and detect the east-west component of the tilt only, they may underestimate the 3D tilt angle of the field.\footnote{By assumption, the usual derivation of radial photospheric field data from LOS images likewise applies only angles in the direction ${\bf\nabla}\rho$ of increasing $\rho$ and not the tilt in the direction perpendicular to ${\bf\nabla}\rho$.}  A purely radial field would give a symmetric V-shaped curve with maximum strength at disk-center and passing through zero at the limbs, while a purely azimuthal field would have an anti-symmetric graph, passing through zero at disk-center and having equal and opposite value at the limbs.  In our example, therefore, the photospheric field seems to be almost radially-directed whereas the chromospheric field has a marked azimuthal signature.   We rejected cases where the graphs were not characterized by simple curves, e.g., cases with fewer than three points and cases failing a reduced $\chi^2$ test.  Some fields do not conform well to a linear combination of radial and azimuthal vectors either because they evolve significantly during their disk passage or because the field structure is complex and does not have a prevalent field direction.  For example, a circle whose flux content points with approximately equal intensity in all directions would give an approximately unchanging average LOS strength throughout its disk passage.  Such a profile cannot be represented by a linear combination of radial and azimuthal vectors and so a successful curve fit would not be possible.  In the majority of our examples, however, a good fit is available.  The photospheric and chromospheric distributions of these east-west tilt angles are shown in histograms in Figure~\ref{hists} for the photospheric (left) and chromospheric (right) measurements.  It is clear that the photospheric histogram is much thinner than the chromospheric one, with peak at about $0^{\circ}$ tilt and a HWHM about $10^{\circ}$.  The chromospheric histogram peaks within $10^{\circ}$ of zero tilt and has HWHM about $30^{\circ}$, with large wings containing many highly tilted fields.  The best-fitting Gaussians to the histograms had maxima at $1.8^{\circ}$ and $5.5^{\circ}$ and width (standard deviation) $10.8^{\circ}$ and $35^{\circ}$.  The histograms are consistent with the picture of steady flux tubes that are intensely concentrated and oriented nearly vertically by fluid forces and buoyancy at photospheric heights and that are free to expand in all directions and tilt angles in the chromosphere.

The asymmetries in Figure~\ref{aveflux} discussed earlier can also be seen in the offsets of the histogram peaks from zero in Figure~\ref{hists}.  These offsets are positive in both photospheric and chromospheric cases.  In histograms of tilt angles for positive and negative fields separately (not shown) the positive and negative photospheric cases are very similar, while in the chromosphere most negative fields have positive tilt and vice versa.  Thus both cases are consistent with a negative (west-east) azimuthal field as seen in Figure~\ref{aveflux}.  In both atmospheric layers, the negative fields outnumber positive fields, biasing the histograms towards positive tilts.

Figure~\ref{qualscatter} shows scatter plots of the disk-center field strength versus east-west tilt angle for the photospheric and chromospheric data sets.  The greater concentration of small tilt angles in the photosphere is again apparent in these plots.  In addition, 66~\% of photospheric data sets with more than three data points are coherent enough for a satisfactory curve fit and tilt angle definition to be possible, compared to 59~\% for the chromospheric data sets.  The vast majority of data sets with no satisfactorily fitted curve, photospheric and chromospheric, were cases with weak field.  The average disk-center field strength in coherent photospheric/chromospheric data sets is 4.9~G/4.4~G compared to 0.69~G/1.30~G for incoherent data sets.  The simplest explanation of this is that these weak fields evolved quickly and randomly as they traversed the disk.  Another striking feature of these plots is the lack of a highly tilted field ($>45^{\circ}$ east or west) in either atmospheric layer with average strength greater than a few tens of Gauss.

\section{Polar fields}
\label{polarfields}

The SOLIS data allow us to extend the analysis of solar poles in several ways.  The sensitivity of the VSM and the high spatial resolution of the images allow us to sample the fields with better spatial resolution than has been possible in the past.  The use of a chromospheric line allows us to estimate the polar field at chromospheric heights for the first time.  Finally, the full-disk vector field measurements produced by the VSM equip us with routine direct measurements of field components perpendicular to the line of sight for the first time.  During the past few years, the SOT-SP instrument on Hinode has
periodically produced vector fields of the polar regions, showing many vertically-oriented intense flux tubes as well as ubiquitous horizontal fields (Tsuneta et al.~2008).

Svalgaard et al.~(1978) inferred the distribution of the polar magnetic flux from the changing orientation of the Sun's rotation axis using Wilcox data.  They compared LOS field strengths at different apertures of their instrument and then fitted a simple forward model for a flux distribution of the form $\cos^n\theta$.  Their best-fitting solution was a radial field with $n=8$.  Using potential coronal field models based on Wilcox and Mt. Wilson data, Wang \& Sheeley~(1988) found a similar cosine colatitude distribution with $n=7.75$ and strength 5.7~G, a similar value to that estimated by Svalgaard et al.  By forward modeling coronal hole sizes and distributions around sunspot minimum using potential field models, Nash et al.~(1988) and Sheeley et al.~(1989) inferred a similar cosine distribution with $n$ taking values in the range $6-8$.

\clearpage

\begin{figure*}[ht]
\begin{center}
\resizebox{0.89\hsize}{!}{\includegraphics*{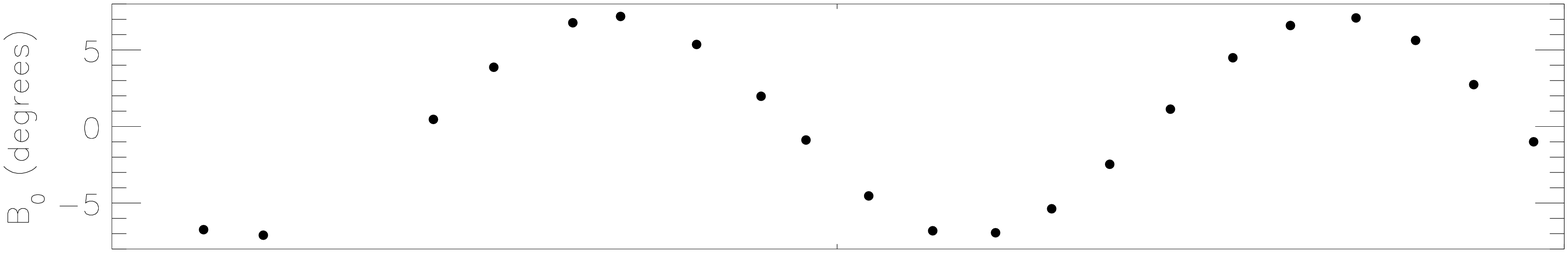}}
\resizebox{0.89\hsize}{!}{\includegraphics*{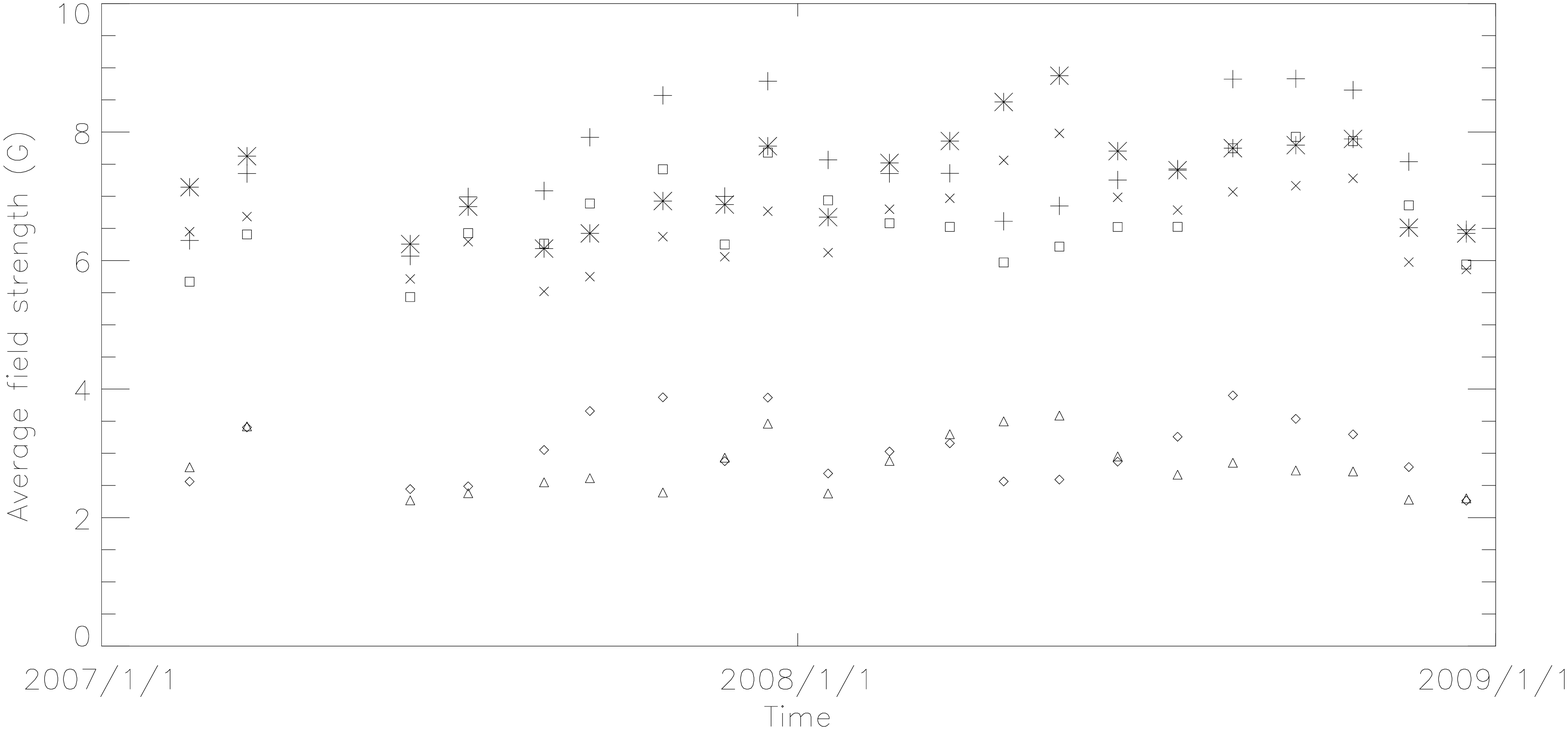}}
\resizebox{0.89\hsize}{!}{\includegraphics*{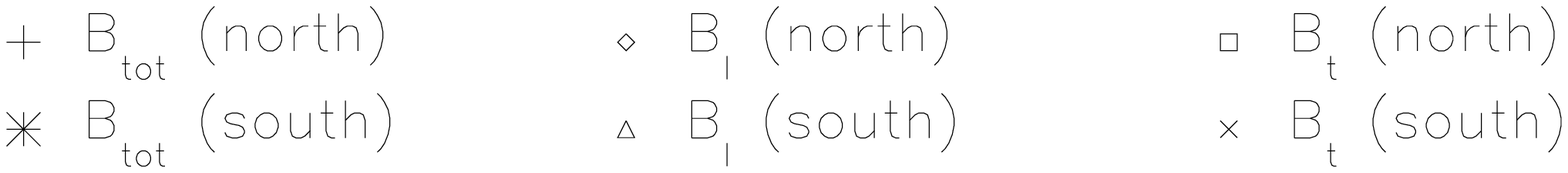}}
\end{center}
\caption{The main panel shows the temporal and latitudinal variation of LOS and transverse fields during 2007 and 2008.  The solar rotation axis tilt angle $B_0$ is indicated by solid circles in the top panel.}
\label{polarvarvectrans}
\end{figure*}

%Figure~\ref{polarvarvec} shows the monthly average LOS fields derived from all 2007 and 2008 full-disk vector magnetograms, averaged over the large dashed circle in Figure~\ref{fdcircles}.  In contrast to the results for the scalar magnetograms above, the lower left panel of Figure~\ref{polarvarvec} shows a net positive field in the northern hemisphere which becomes close to zero at high latitudes.  The southern measurements are more scattered.

\begin{figure*}[ht]
\begin{center}
\resizebox{0.69\hsize}{!}{\includegraphics*{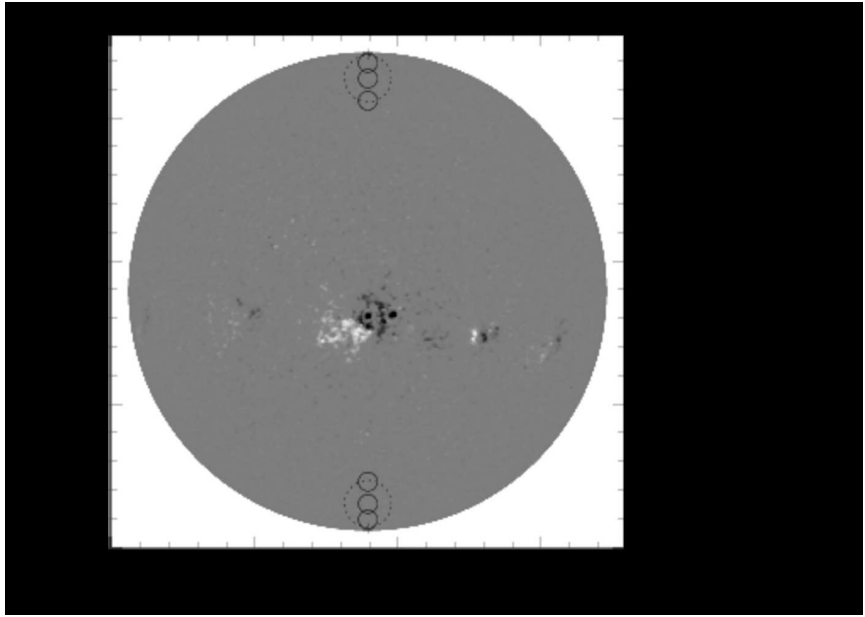}}
\end{center}
\caption{A full-disk sky image of the Sun with circles marked in the northern/southern hemisphere whose centers coincide with solar latitudes $\pm 60^{\circ}$, $\pm 70^{\circ}$ and $\pm 80^{\circ}$ when the rotation axis tilt attains extrema $\pm 7.25^{\circ}$.   The + symbols mark the furthest extent of the poles onto the disk at these rotation axis tilt extrema.}
\label{fdcircles}
\end{figure*}

Figure~\ref{polarvarvectrans} shows monthly average unsigned LOS and transverse field components.  These are averaged over the large dashed circles over-plotted on the full-disk image in Figure~\ref{fdcircles}.  Whereas in the earlier sections we worked with circles confined to fixed latitudes within remapped images, here we instead consider circles in full-disk image coordinates whose positions are fixed at at certain positions on the disk, and whose heliographic latitudes therefore vary as the Sun tilts back and forth with respect to our line of sight.  We found that this approach worked best at polar latitudes because circles contain equal numbers of pixels at all times giving each time series more coherence than with alternative approaches.  The average latitudes of the measurement locations within these large circles is $\pm 70^{\circ}$ when the rotation axis tilt angle attains its extrema $B_0=\pm 7.25^{\circ}$.  As this tilt angle changes with period a year, the heliographic latitudes of the measurements change correspondingly.   Averages are calculated from all 2007 and 2008 quick-look full-disk vector magnetograms.  The transverse measurements have a $180^{\circ}$-degree ambiguity and so their unsigned strengths are plotted, as are the unsigned strengths of the LOS measurements.  For an approximately axisymmetric, static polar flux distribution the LOS measurements at fixed position on the disk should display an annual periodicity.  Assuming that the polar field is indeed quasi-static and nearly axisymmetric, it is evident from the plot that these measurements are not sensitive enough to show the expected sinusoidal dependence on time.  However, the difference in strength between the LOS and transverse field components is clear.  The ratio between transverse and LOS components is approximately $\tan (70^{\circ} )$, which is consistent with a radial field.  The results shown here derive from simple inversions based on a weak-field approximation and therefore are strictly preliminary.  The vector measurements are currently being fully inverted.  A future study will investigate the properties of more precise, fully inverted measurements of polar fields.  In the meantime, we focus on the polar measurements in the scalar LOS images discussed in the previous two sections.

\clearpage

\clearpage

\begin{figure*}[ht]
\begin{center}
\resizebox{0.79\hsize}{!}{\includegraphics*{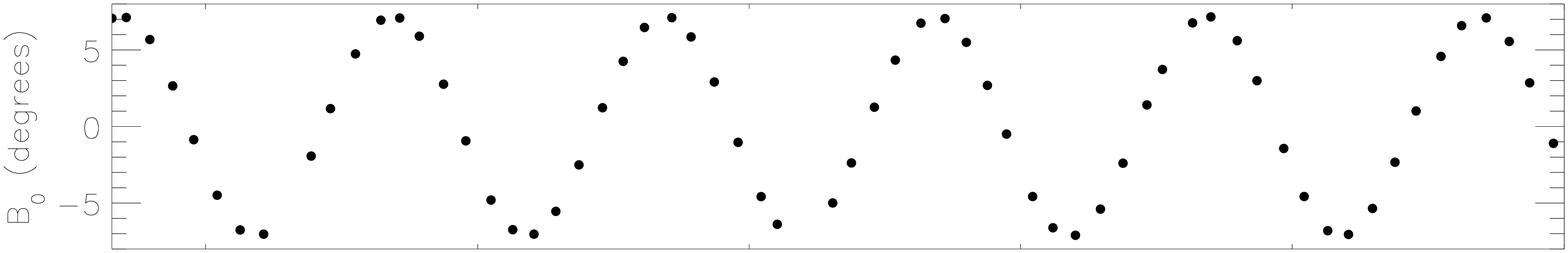}}
\resizebox{0.79\hsize}{!}{\includegraphics*{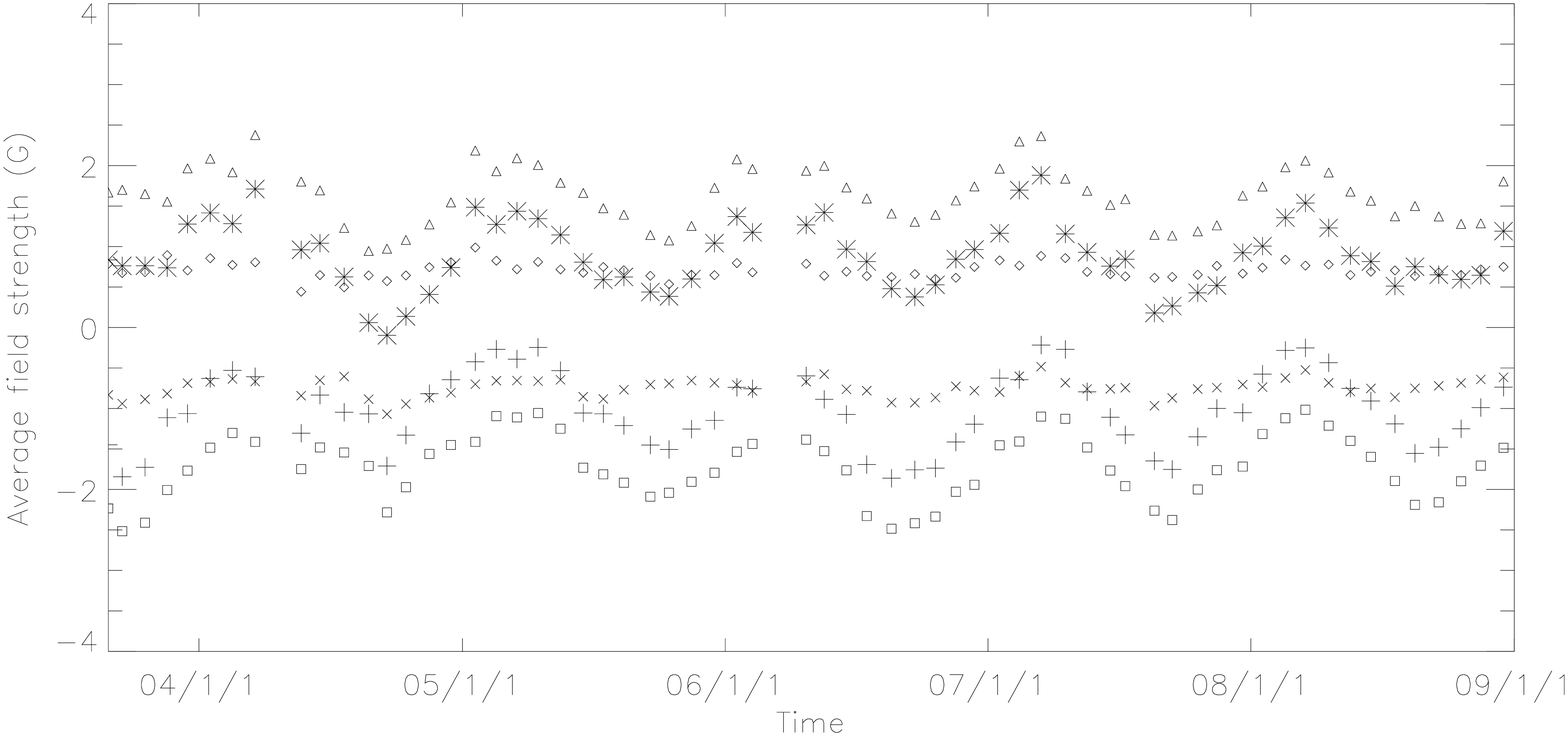}}
\resizebox{0.79\hsize}{!}{\includegraphics*{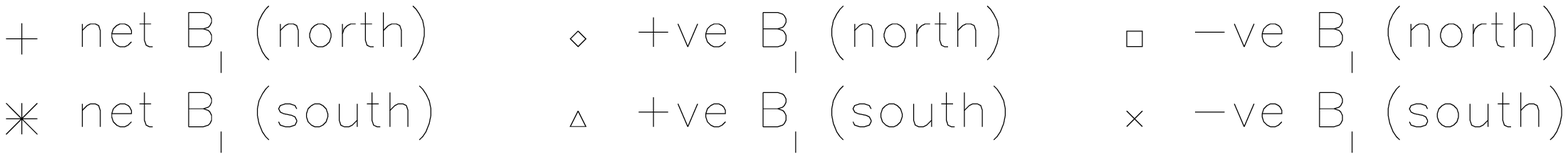}}
\end{center}
\caption{Temporal variation (main panel) of the northern and southern polar photospheric magnetic fields averaged month-to-month within the solid circles of Figure~\ref{fdcircles} corresponding to $\pm 60^{\circ}$ latitude.  The solar rotation axis tilt angle $B_0$ is indicated by solid circles in the top panel.}
\label{polarvarphot70}
\end{figure*}

Figure~\ref{polarvarphot70} shows the monthly average photospheric field variation within a north/south pair of circles in Figure~\ref{fdcircles} whose average latitudes are $\pm 70^{\circ}$ at tilt angle extrema $B_0=\pm 7.25^{\circ}$.  The annual periodicity in the measurements is obvious.  If the polar field were uniform and strictly radial, the measurements of their LOS component would be approximately constant as the rotation axis tips back and forth between $B_0=\pm 7.25^{\circ}$.  Figure~\ref{polarvarphot70} shows that these measurements do change significantly.  Of course, the latitude of our chosen portion of the disk changes with annual periodicity also.  

\clearpage

\begin{figure*}[ht]
\begin{center}
\resizebox{0.345\hsize}{!}{\includegraphics*{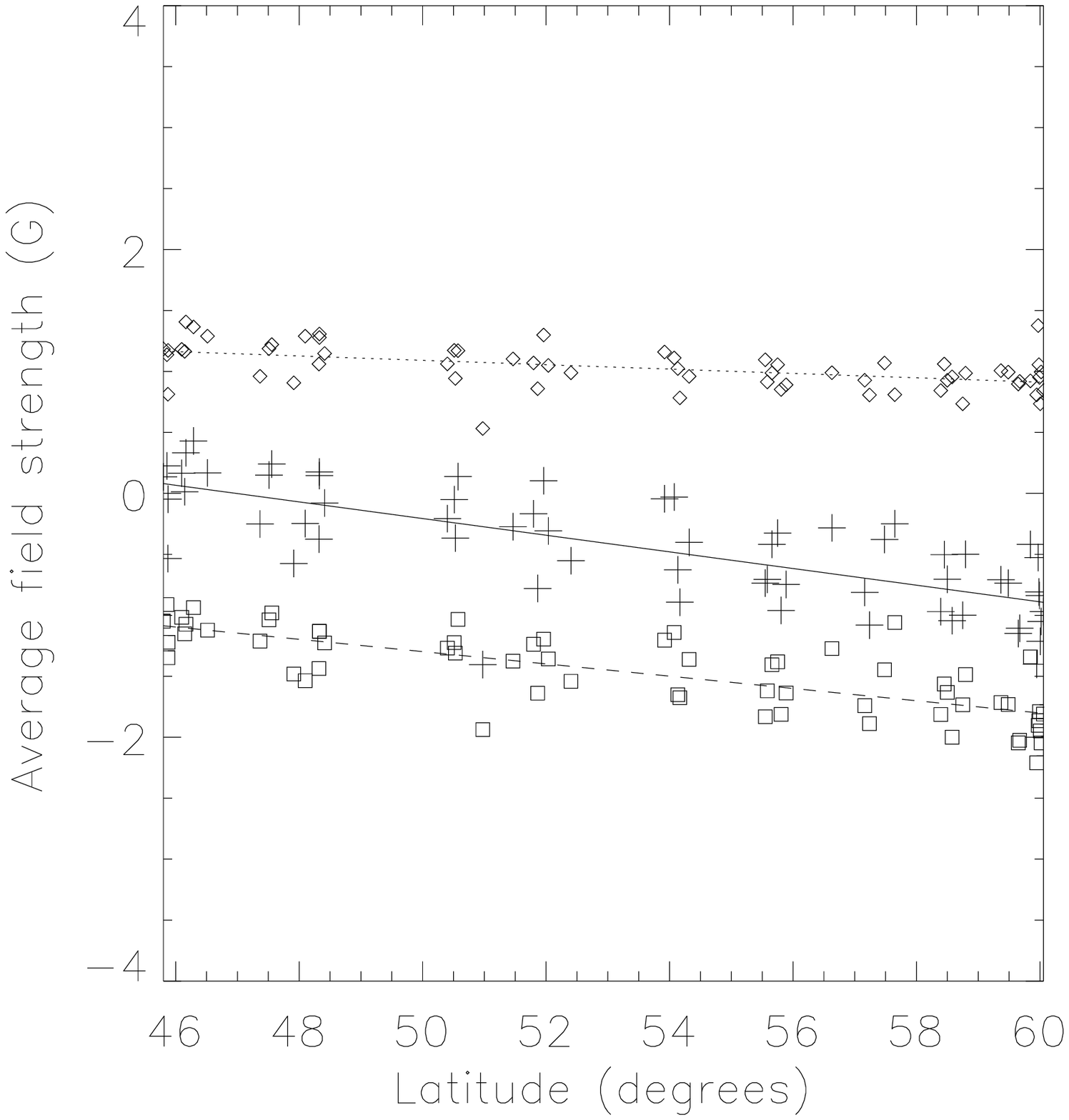}}
\resizebox{0.345\hsize}{!}{\includegraphics*{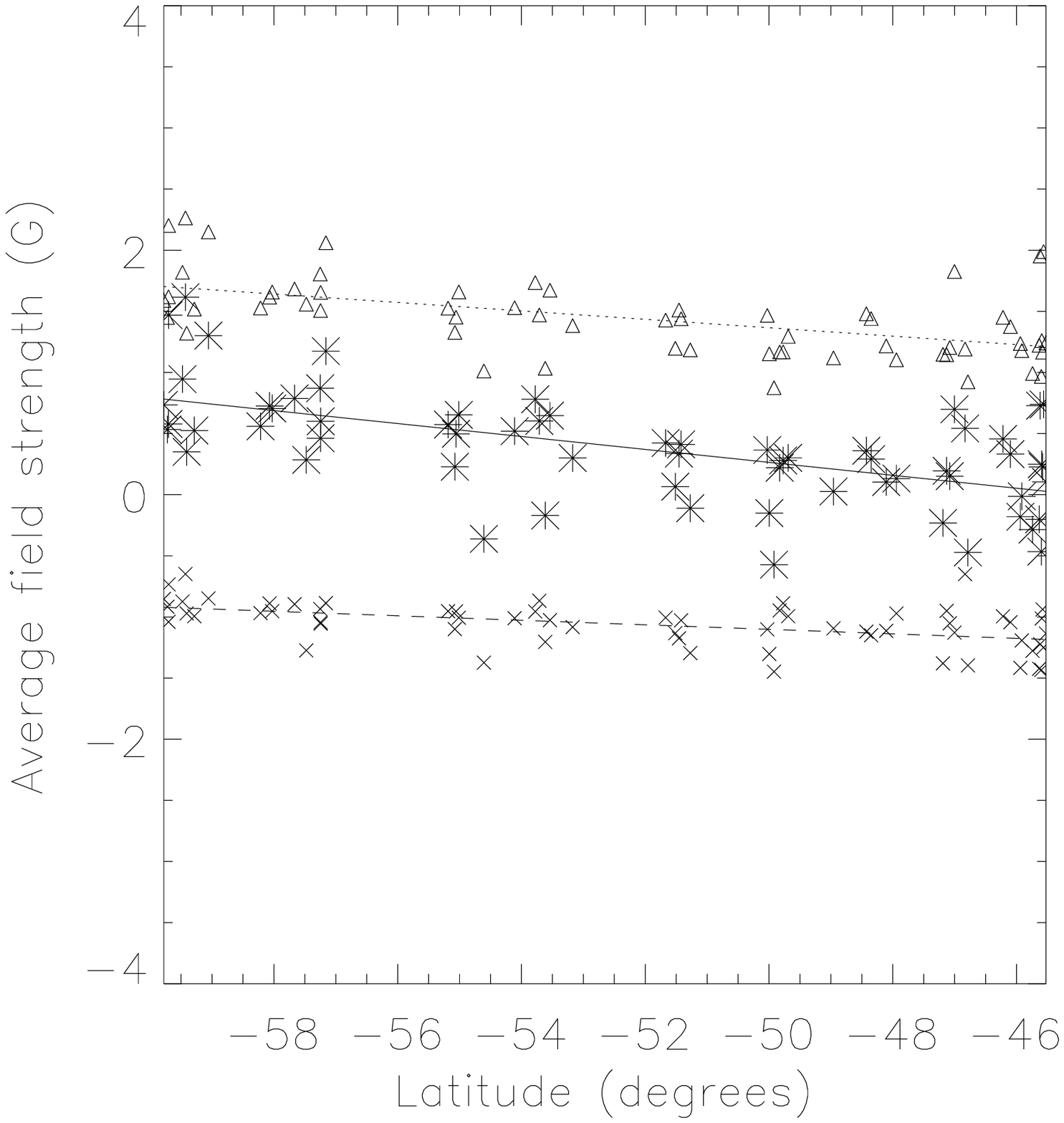}}
\resizebox{0.345\hsize}{!}{\includegraphics*{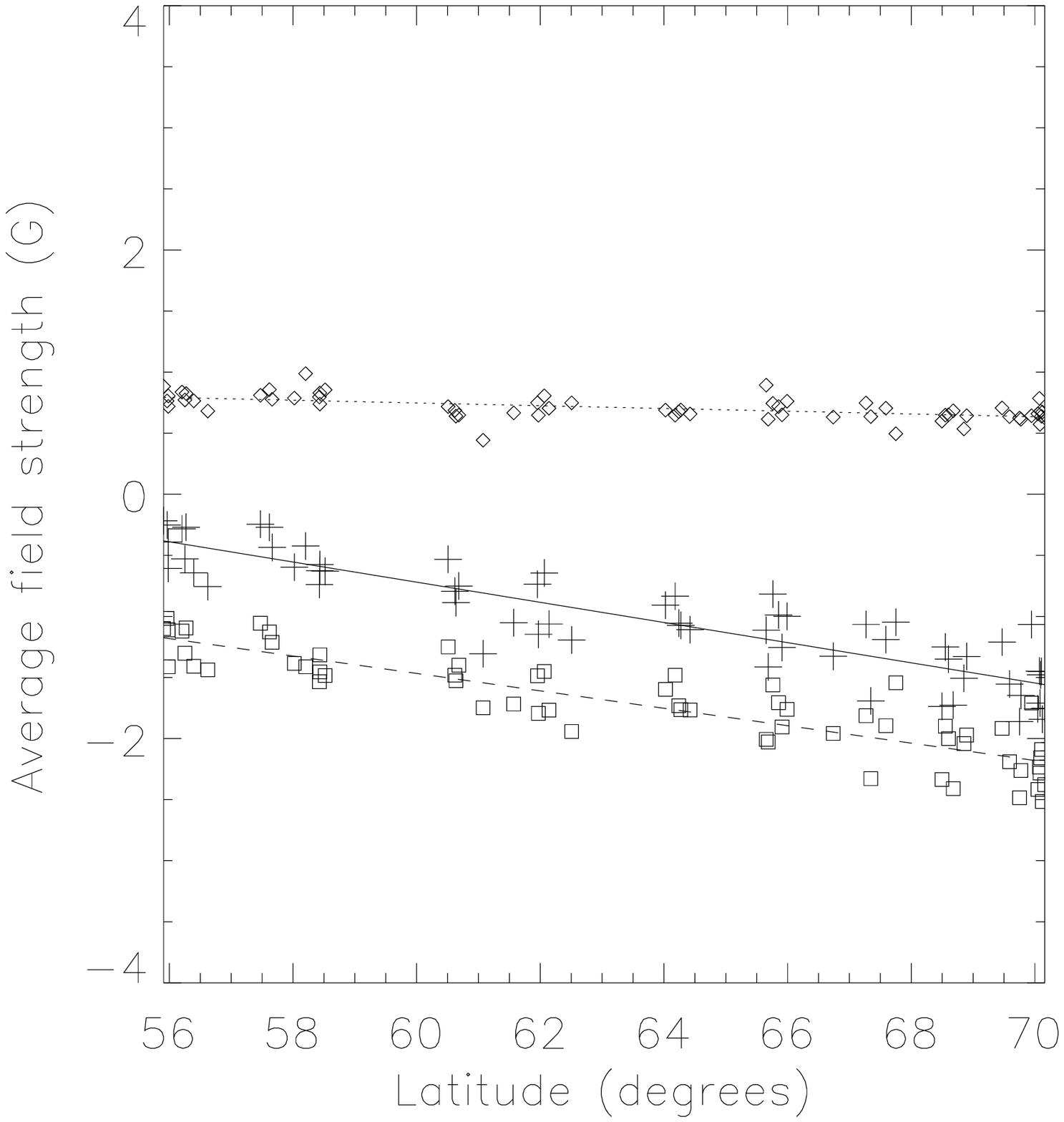}}
\resizebox{0.345\hsize}{!}{\includegraphics*{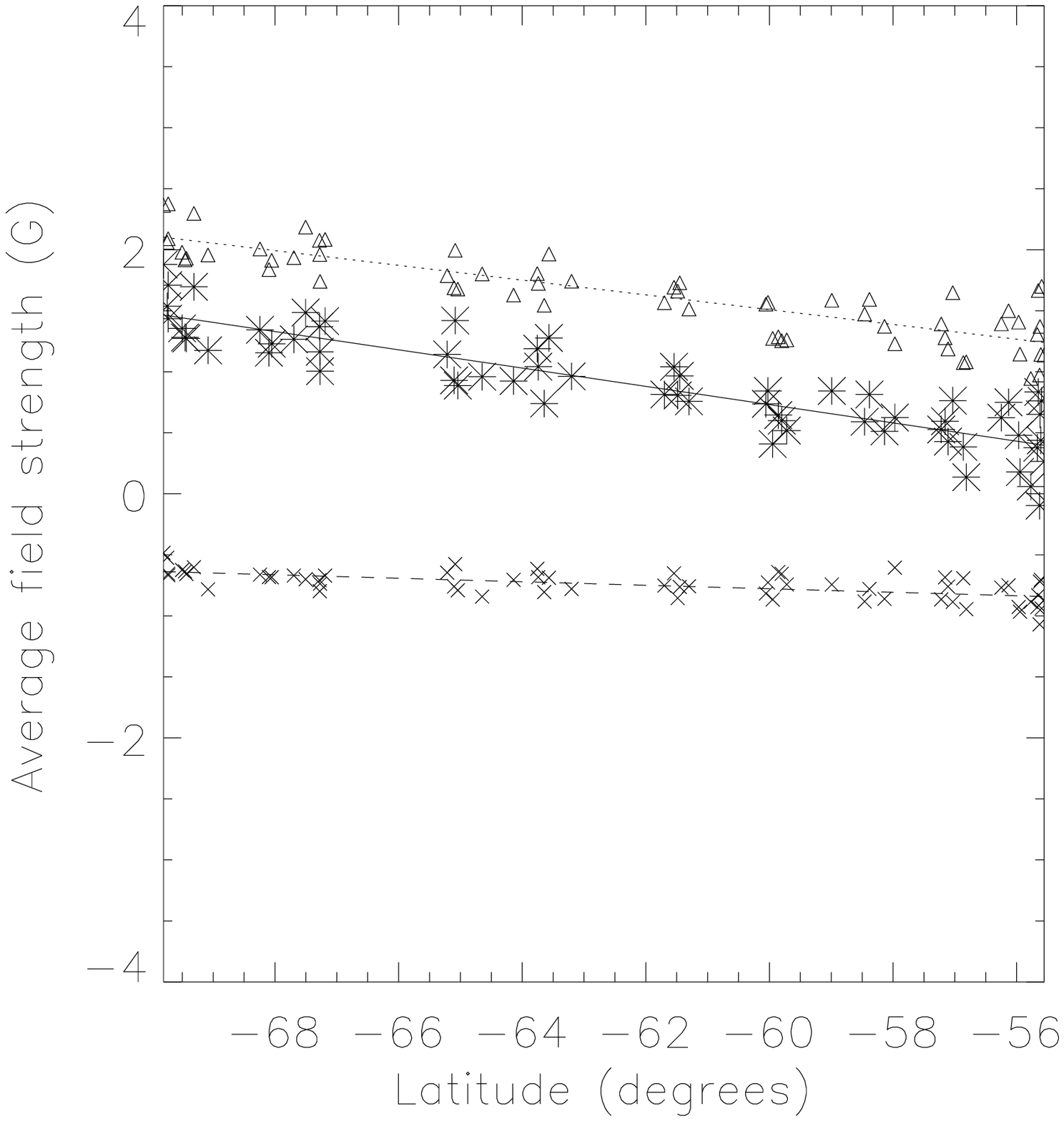}}
\resizebox{0.345\hsize}{!}{\includegraphics*{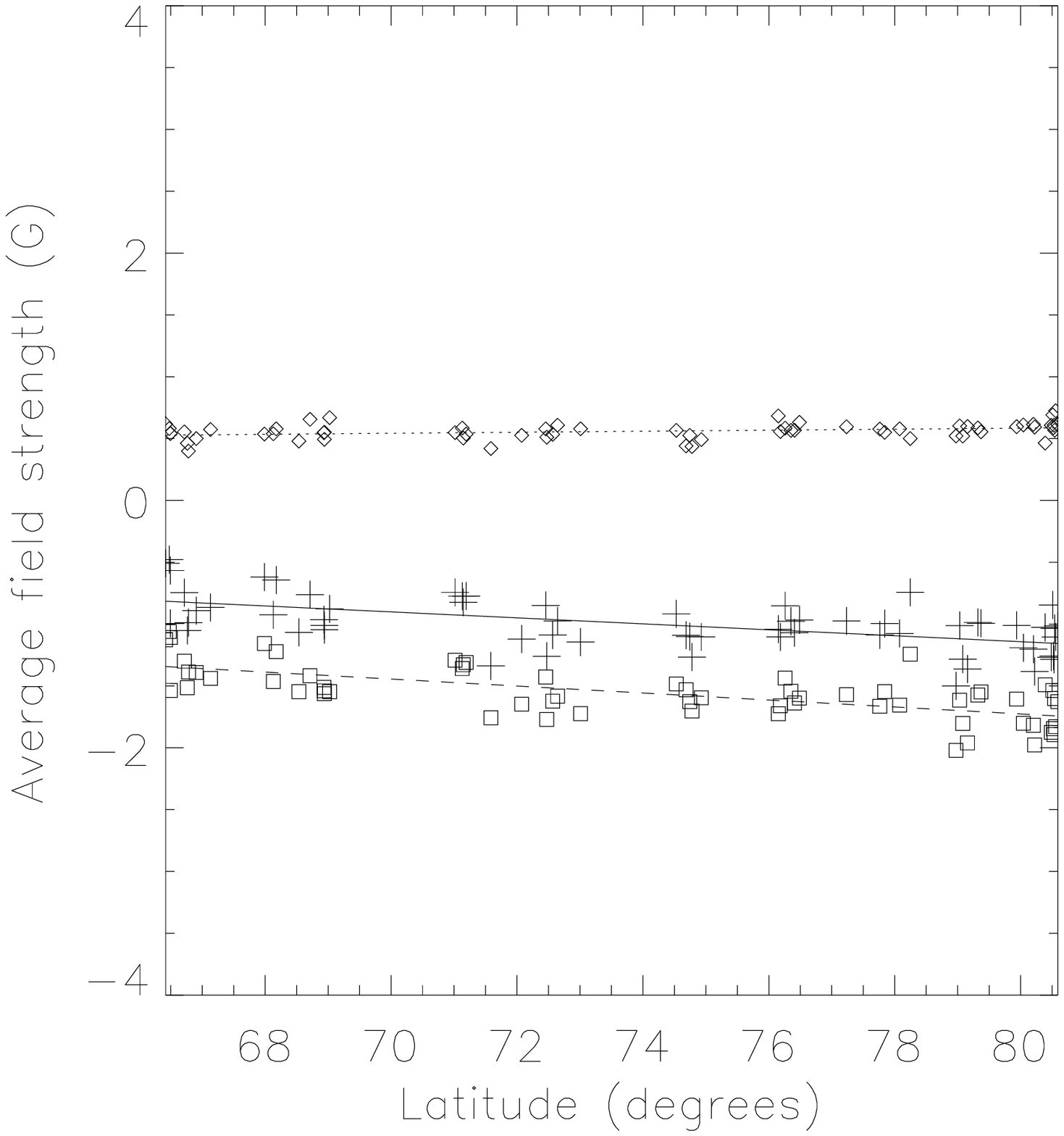}}
\resizebox{0.345\hsize}{!}{\includegraphics*{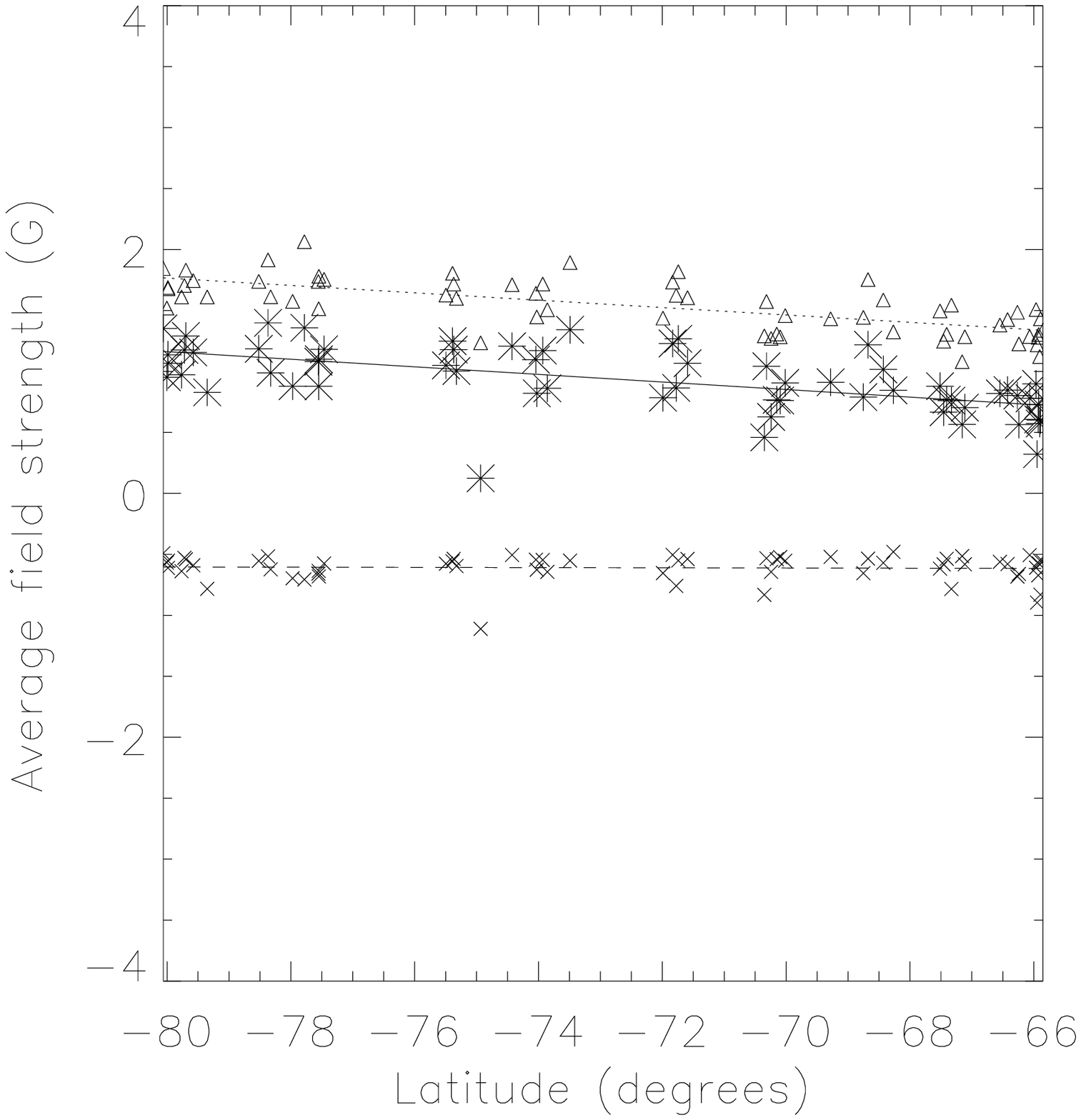}}
\resizebox{0.69\hsize}{!}{\includegraphics*{f11c.eps}}
\end{center}
\caption{Latitudinal variation of the northern (left pictures) and southern (right pictures)  photospheric polar magnetic fields averaged month-to-month within the solid circles of Figure~\ref{fdcircles} corresponding to $\pm 60^{\circ}$ (top pictures), $\pm 70^{\circ}$ (middle pictures) and $\pm 80^{\circ}$ (bottom pictures) latitude.  The solid, dotted and dashed lines show the best linear fits to the net, positive and negative field data, respectively.}
\label{polarvarphotlats}
\end{figure*}

Figure~\ref{polarvarphotlats} shows the photospheric field variation within the three north/south pairs of circles shown in  Figure~\ref{fdcircles} as functions of latitude.  Since these circles are fixed at constant position on the disk $\rho$, their latitude varies sinusoidally over a range of $14.5^{\circ}$ as the tilt angle $B_0$ of the Sun's rotation axis oscillates between $\pm 7.25^{\circ}$ annually.  For a quasi-static, nearly axisymmetric polar field we therefore expect the LOS field in a given circle to take approximately the same value every time that circle revisits a given latitude.  We therefore expect the LOS field to be a well-defined function of latitude for each circle.   Figure~\ref{polarvarphotlats} shows that, in each case, the dependence of the measured LOS field on latitude is indeed simple and well-defined.  In each row of Figure~\ref{polarvarphotlats},  the heliocentric angle is the same for every data point: $60^{\circ}$, $70^{\circ}$ and $80^{\circ}$, respectively.  These figures show that the LOS components of the fields increase in strength towards the poles, implying that either (i) the total field strength must also increase towards the poles, or (ii) the fields nearer the pole are more tilted towards the LOS direction.  We will investigate which scenario is the correct one presently.  In the middle plots, both northern and southern fields are distinctly polarized.  There is a pronounced latitude dependence in the northern negative field measurements such that the negative and net fields increase in strength towards the pole.  There is a corresponding dependence in southern positive measurements, albeit to a lesser degree.  The opposite-polarity field measurements hardly vary in strength at the high-latitude measurements in the bottom plots, and yet they are almost as strong as the poles' dominant polarities.  The strong $B$-angle dependence of the polar polarities shows through in the net field measurements.

\clearpage

\begin{figure*}[ht]
\begin{center}
\resizebox{0.395\hsize}{!}{\includegraphics*{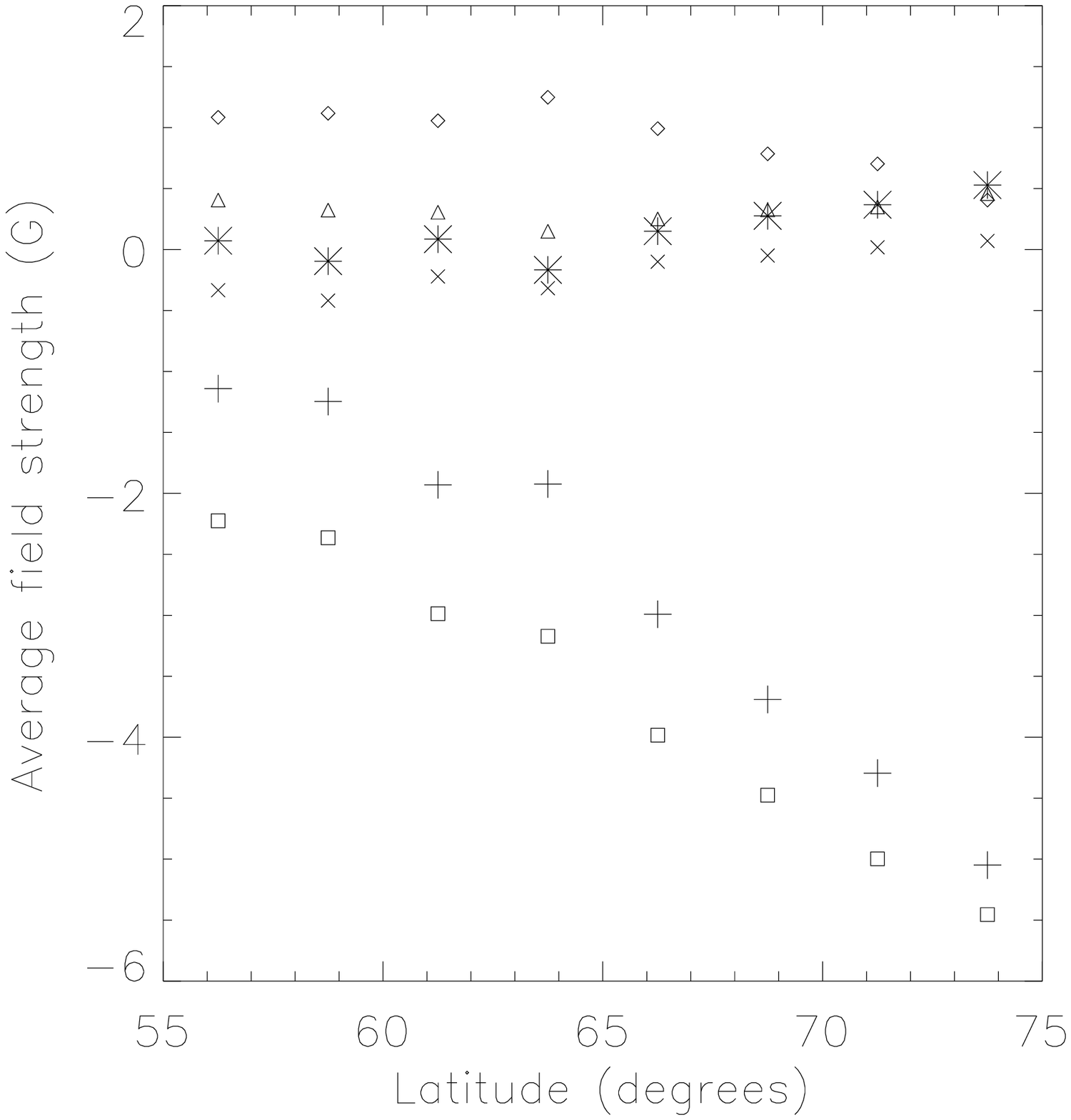}}
\resizebox{0.395\hsize}{!}{\includegraphics*{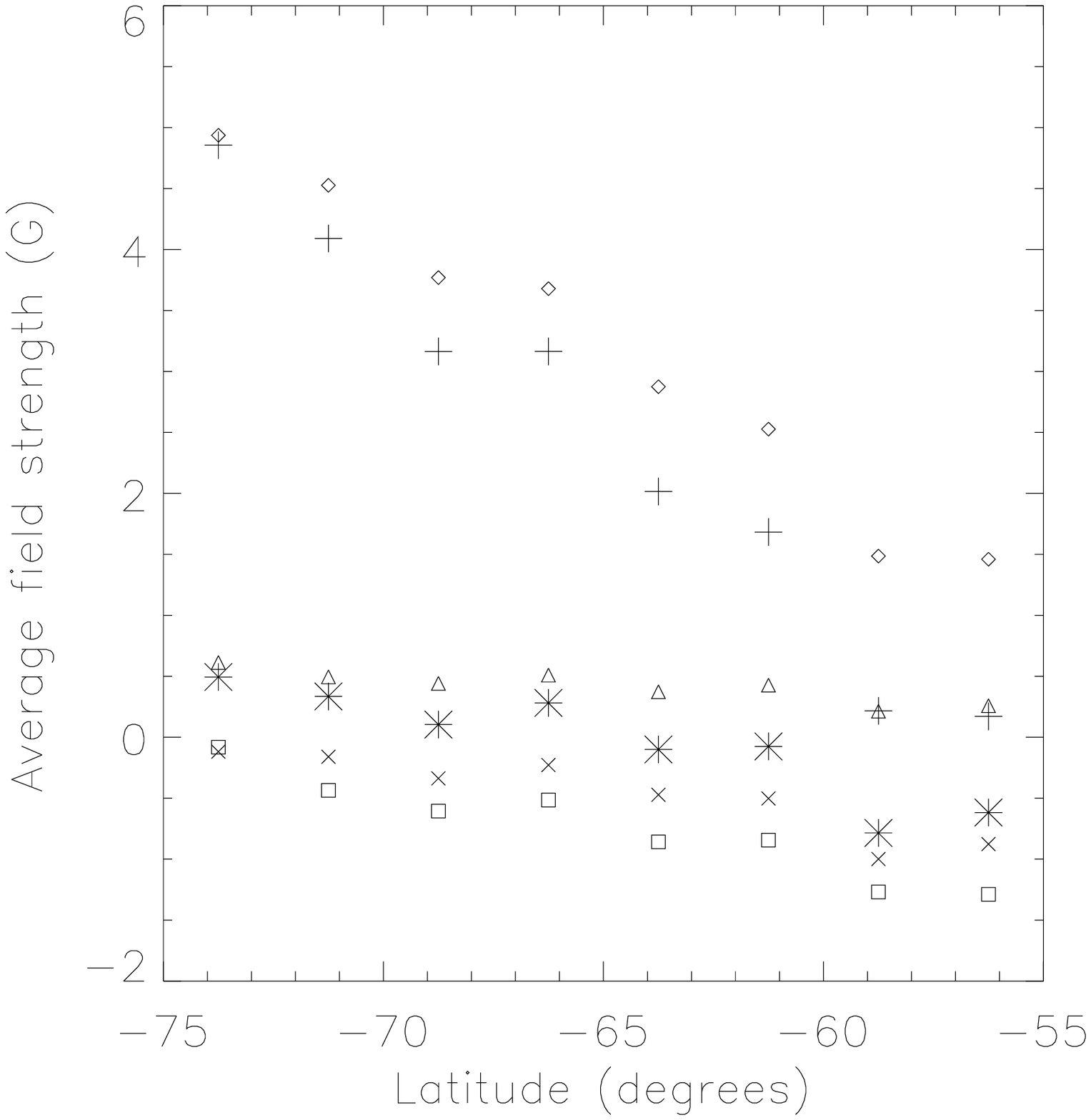}}
\resizebox{0.79\hsize}{!}{\includegraphics*{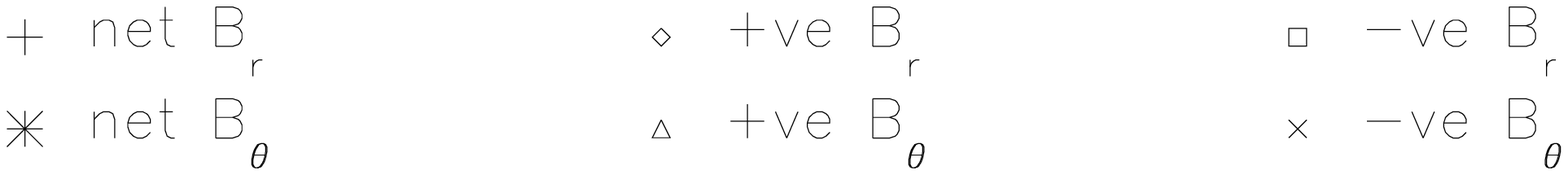}}
\end{center}
\caption{The estimated radial and poloidal photospheric field components as functions of latitude during 2003-2008 for the north pole (left) and the south pole (right).  Note the domination of the radial component at both poles beyond $60^{\circ}$.}
\label{poletilts}
\end{figure*}

We now decompose the polar fields into radial and poloidal components.  Because of the small tilt angle of the solar rotation axis relative to the ecliptic plane, we cannot perform this decomposition exactly as we did with the near-equatorial fields earlier.  Figure~\ref{polarvarphotlats} shows the line-of-sight fields at common latitudes as diagnosed at different locations on the solar disk.  We proceed by assuming that the two LOS measurements at different times of year at different viewing angles $\rho$ but at the same latitude represent different projections of the same quasi-static magnetic vector.  We omit the azimuthal component $B_{\phi}$ because it is not detected at the central meridian.  By comparing LOS measurements deriving from a single latitude but from different locations on the solar disk, both on the central meridian but with different heliocentric angles $\rho_1\neq\rho_2$, the field vector components  $B_r$ and $B_{\theta}$ at this latitude can be deduced.  Using linear coefficients $c_1$, $c_2$, $c_3$ and $c_4$ to describe two straight lines sharing a common latitude $L$,  such as those in Figures~\ref{polarvarphotlats}, we solve the system of equations,

\begin{eqnarray}
B_r\cos\rho_1 + B_{\theta}\sin\rho_1 & = & c_1 + c_2 L,\\
B_r\cos\rho_2 + B_{\theta}\sin\rho_2 & = & c_3 + c_4 L, \label{inversioneqs}
\end{eqnarray}

\noindent
to estimate the magnetic vector $(B_r, B_{\theta} )$ at latitude $L$.   Thus, we estimated the radial and poloidal fields at selected latitudes.  Figure~\ref{poletilts} shows the results.  Between about $\pm 55^{\circ}$ and $\pm 75^{\circ}$ in the photosphere the polar cap appears to begin at about $60^{\circ}$.  From there, the field strength increases monotonically and the field becomes increasingly unipolar as one moves poleward until about $75^{\circ}$ where the results become unreliable.  Furthermore, the radial component is significantly stronger than the poloidal component in the polar cap, indicating that the photospheric is approximately radial at these latitudes.  For estimates beyond $\pm 75^{\circ}$, the assumptions used here break down, fewer field measurements contribute to each data point and so the results become noisier.  Below $60^{\circ}$ the fields are weak, and the method does not determine a preferred field direction there.

\clearpage

\begin{figure*}[ht]
\begin{center}
\resizebox{0.79\hsize}{!}{\includegraphics*{f11a.eps}}
\resizebox{0.79\hsize}{!}{\includegraphics*{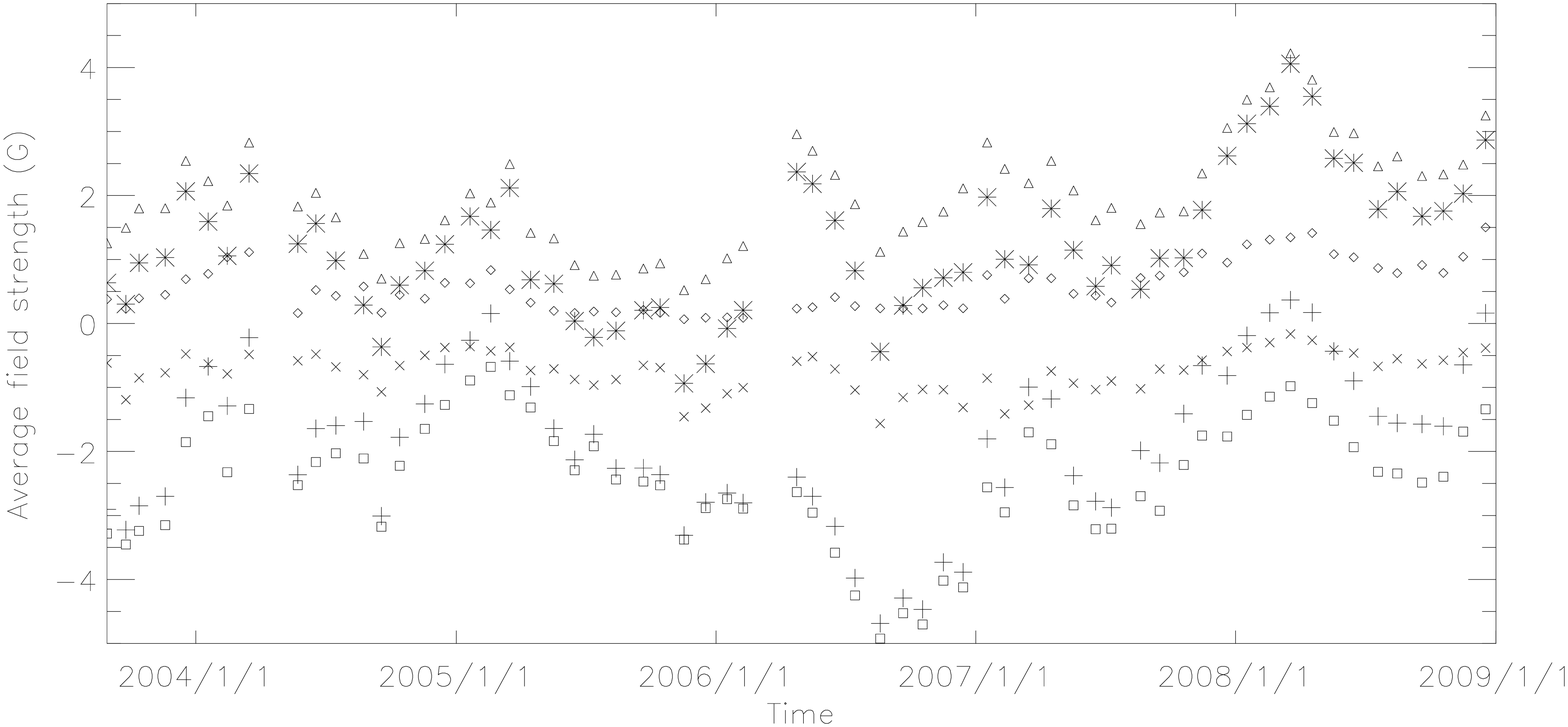}}
\resizebox{0.395\hsize}{!}{\includegraphics*{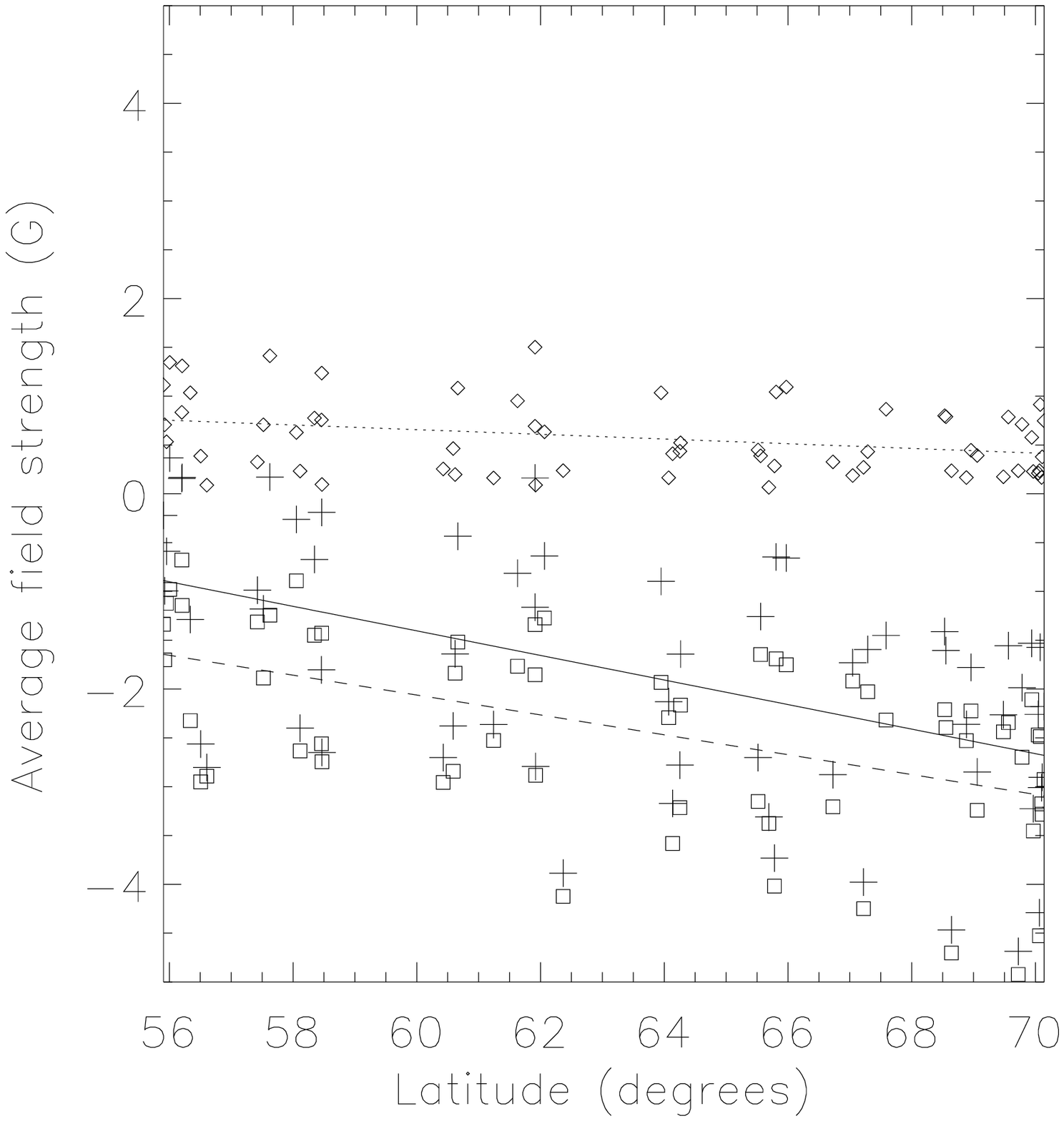}}
\resizebox{0.395\hsize}{!}{\includegraphics*{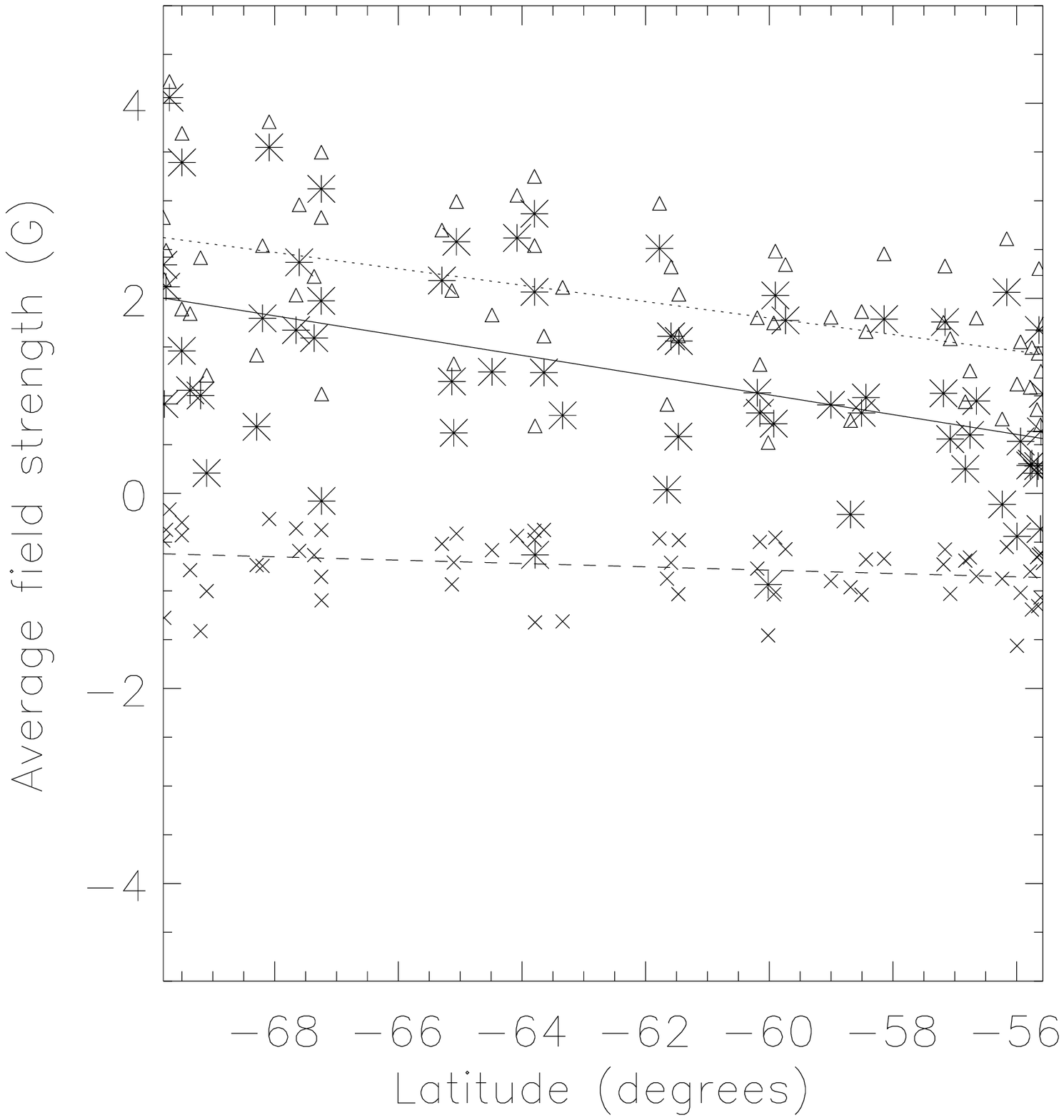}}
\resizebox{0.79\hsize}{!}{\includegraphics*{f11c.eps}}
\end{center}
\caption{The temporal (second panel from top) and latitudinal (bottom panels) variations of chromospheric fields during 2003-2008 within the solid circles of Figure~\ref{fdcircles} corresponding to $\pm 70^{\circ}$.  Note the non-steady nature of the fields' evolution compared to the photospheric fields in, e.g., Figure~\ref{polarvarphot70}.  In the bottom panels the solid, dotted and dashed lines show the best linear fits to the net, positive and negative field data, respectively.   The solar rotation axis tilt angle $B_0$ is indicated by solid circles in the top panel.}
\label{polarvarchromfull}
\end{figure*}

\clearpage

\begin{figure*}[ht]
\begin{center}
\resizebox{0.395\hsize}{!}{\includegraphics*{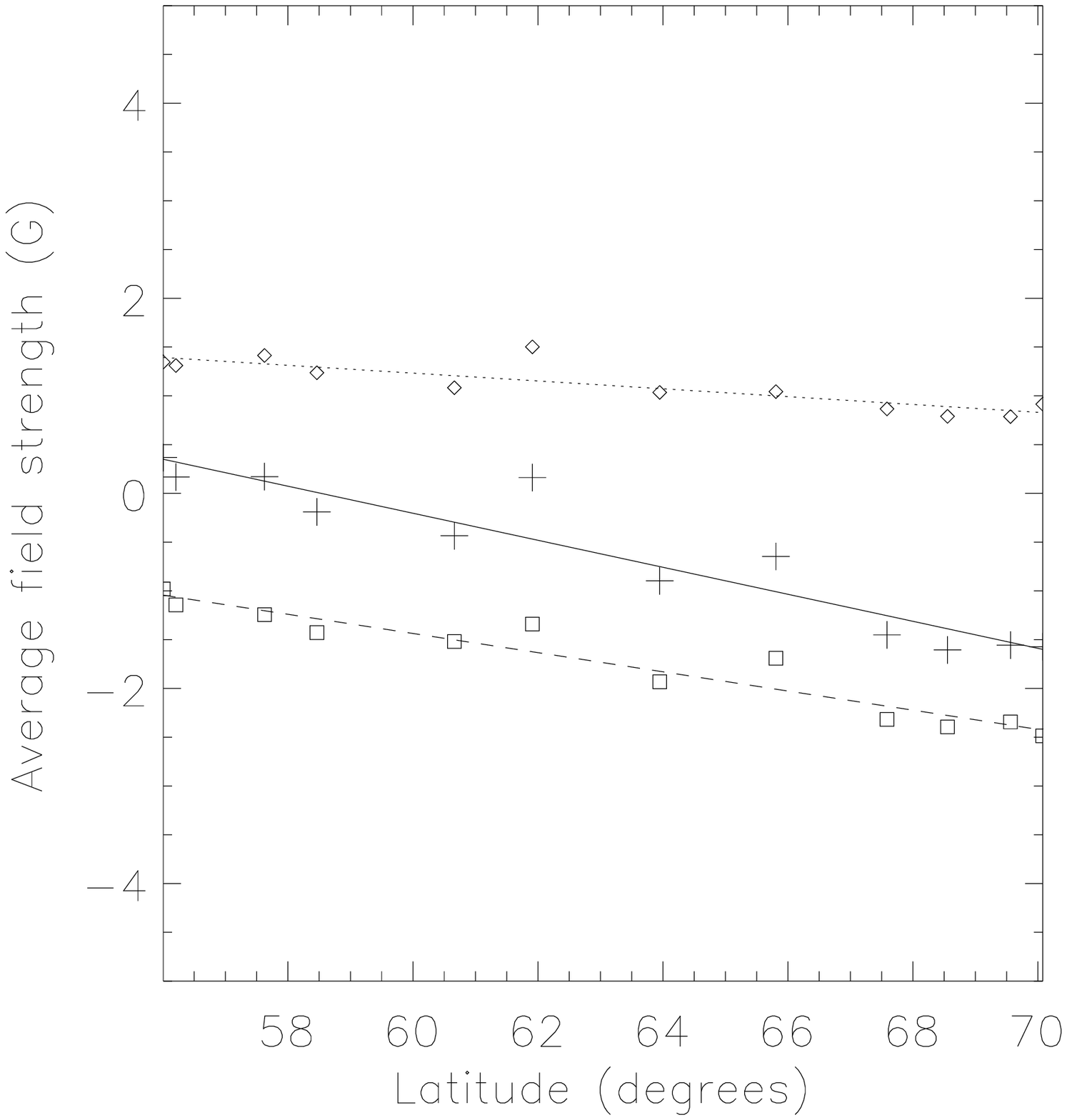}}
\resizebox{0.395\hsize}{!}{\includegraphics*{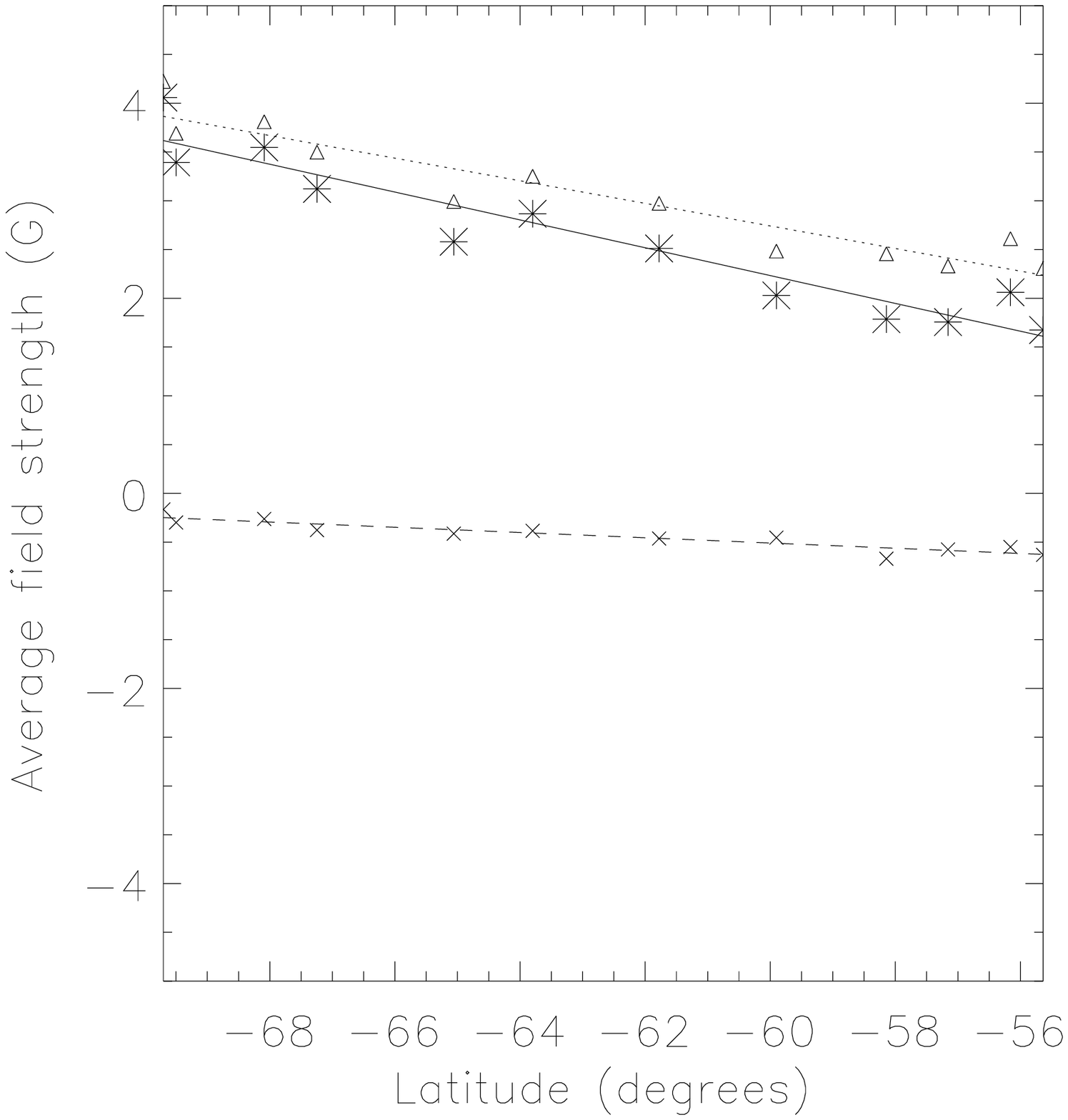}}
\resizebox{0.79\hsize}{!}{\includegraphics*{f11c.eps}}
\end{center}
\caption{The latitudinal variations of chromospheric fields within the $\pm 70^{\circ}$ circles during 2008 only, throughout which time the fields were steady.  The solid, dotted and dashed lines show the best linear fits to the net, positive and negative field data, respectively.}
\label{polarvarchrom08}
\end{figure*}

\clearpage

\begin{figure*}[ht]
\begin{center}
\resizebox{0.395\hsize}{!}{\includegraphics*{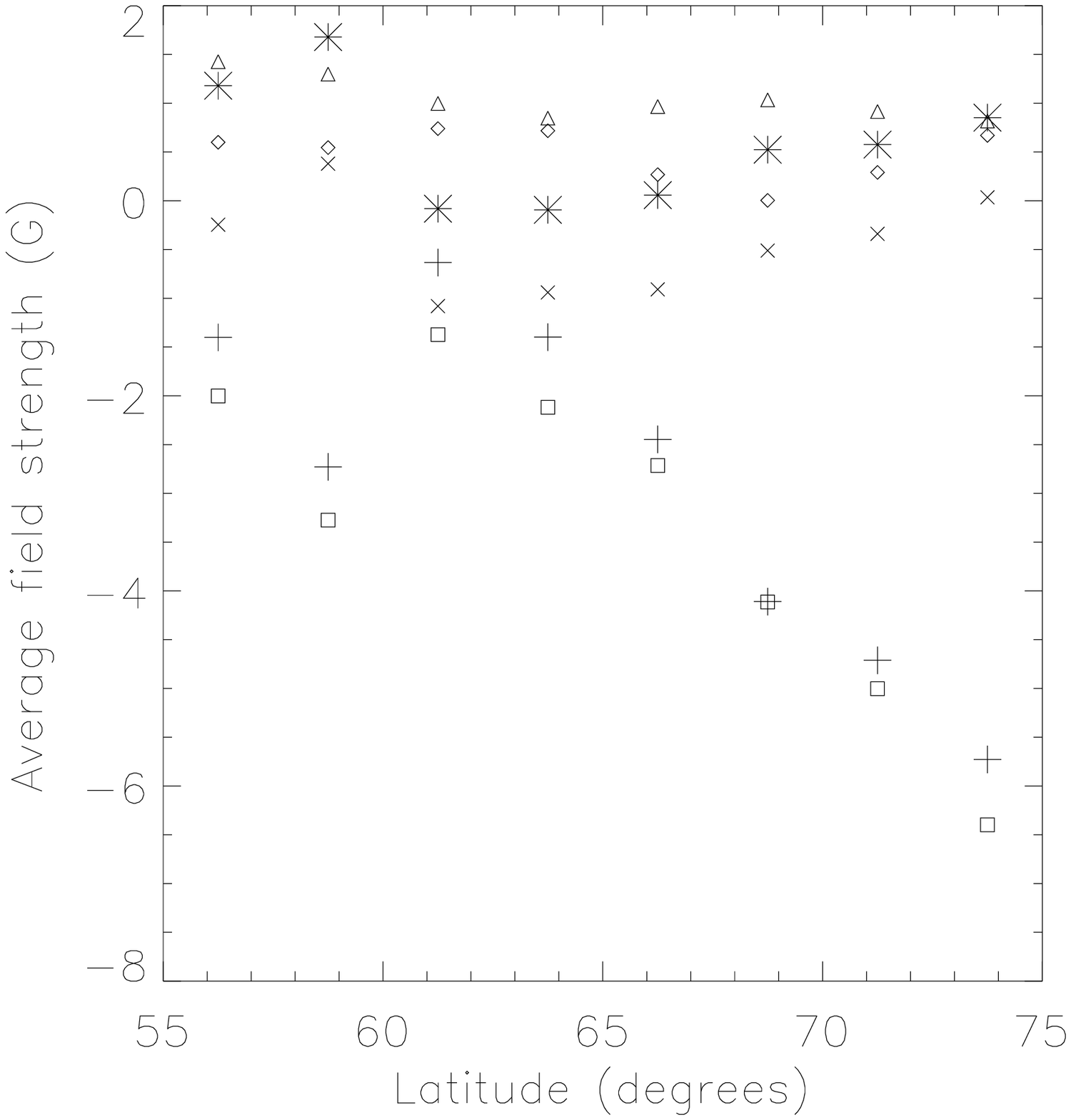}}
\resizebox{0.395\hsize}{!}{\includegraphics*{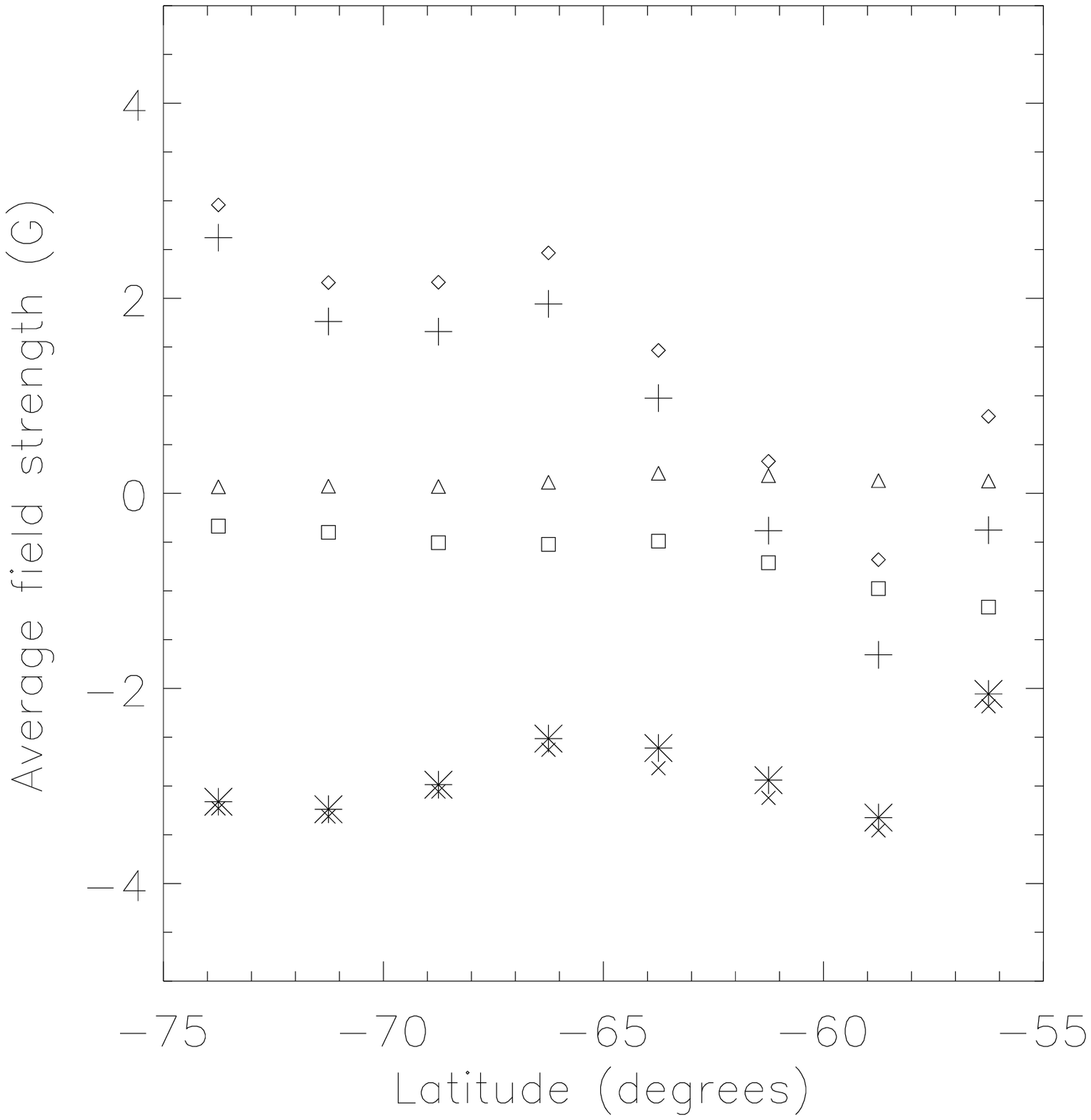}}
\resizebox{0.79\hsize}{!}{\includegraphics*{f13c.eps}}
\end{center}
\caption{The estimated radial and poloidal chromospheric field components as functions of latitude during 2008 for the north pole (left) and the south pole (right).  Note the super-radial field expansion at the south pole.}
\label{poletiltschromo}
\end{figure*}

Figure~\ref{polarvarchromfull} shows the monthly average field variations in the chromosphere, averaged over the solid circles in Figure~\ref{fdcircles} corresponding to latitudes $\pm 70^{\circ}$.  Averaging over smaller circles produces more scattered graphs.  The positive and negative LOS fields have greater maximum strengths than their corresponding photospheric measurements shown in Figure~\ref{polarvarphot70}.  This seems to be an effect of the canopy structure of the field expansion in this atmospheric layer.  Unlike the photospheric cases, the chromospheric time variation seems to show some cycle-dependence.  In the early part of the sample, which corresponds to a relatively active phase of the solar cycle, there is a lot of scatter in the measurements. which appears to be mostly due to complex, highly non-axisymmetric field geometry.  The complex field geometry seems likely to be associated with the influence of strong active fields.  The quality of these chromospheric measurements is comparable to that of the LOS photospheric measurements, but active fields evidently have more influence on the polar field inclination in the chromosphere than in the photosphere.  The later chromospheric measurements show a relatively clean sinusoidal variation suggesting a simple chromospheric field structure during the recent quiet phase of the cycle.  There are large excursions of the chromospheric averages in the north in late 2006 and in the south in early 2008.  There was a small burst of activity around the end of 2006 and the first Cycle-24 fields began to appear in early 2008, but these fields appeared in the southern and northern hemispheres, respectively.  If they are related to the polar excursions reported here, the relationship is not a simple one.  We can at least say that the active and quiet phases of the solar cycle do appear to be reflected in correspondingly complex and simple signals in the polar field measurements.  Comparing the lower panels of Figure~\ref{polarvarchromfull} with Figure~\ref{polarvarchrom08}, which contains only measurements from 2008, emphasizes the simplicity of the field during 2008. 

Figure~\ref{poletiltschromo} shows the radial and poloidal field components during 2008, as determined by the method described above using Equations~(\ref{inversioneqs}).  While the north pole adheres mostly to the pattern described above for the photospheric fields, with the exception of a hint of poleward-tilting field below $60^{\circ}$, the south pole is more complicated.  The  radial component does increase in strength with latitude but the poloidal component is significant for all latitudes.  This is evidence that the south polar field was expanding super-radially (i.e. tilted away from the pole) at chromospheric heights during 2008.  Perhaps this is due to a relatively strong, concentrated south pole whose flux expands into the atmosphere.  Earlier chromospheric data do not show steady enough field patterns for the radial and poloidal components to be determined by these methods.

\clearpage

\begin{figure*}[ht]
\begin{center}
\resizebox{0.69\hsize}{!}{\includegraphics*{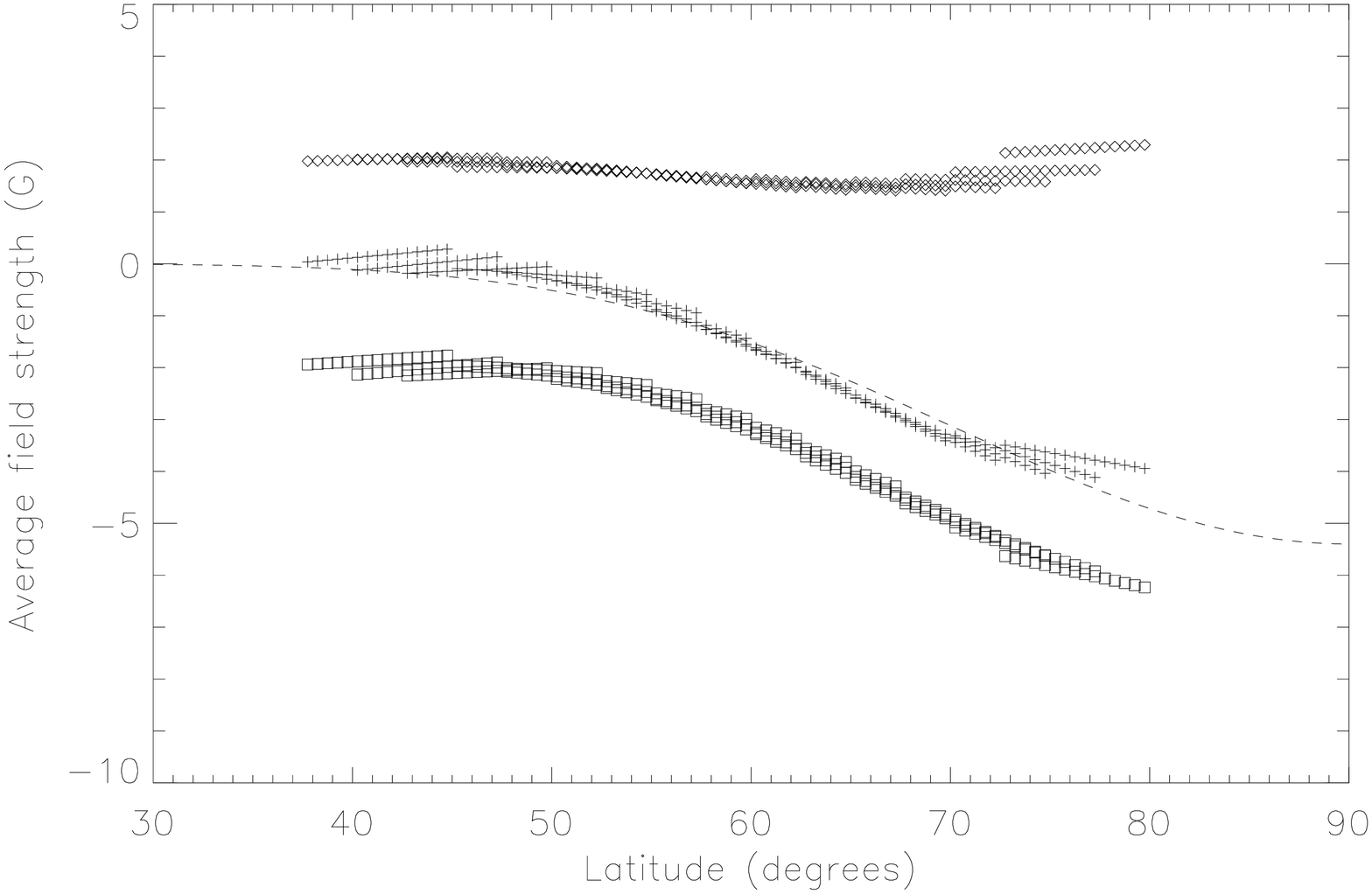}}
\resizebox{0.69\hsize}{!}{\includegraphics*{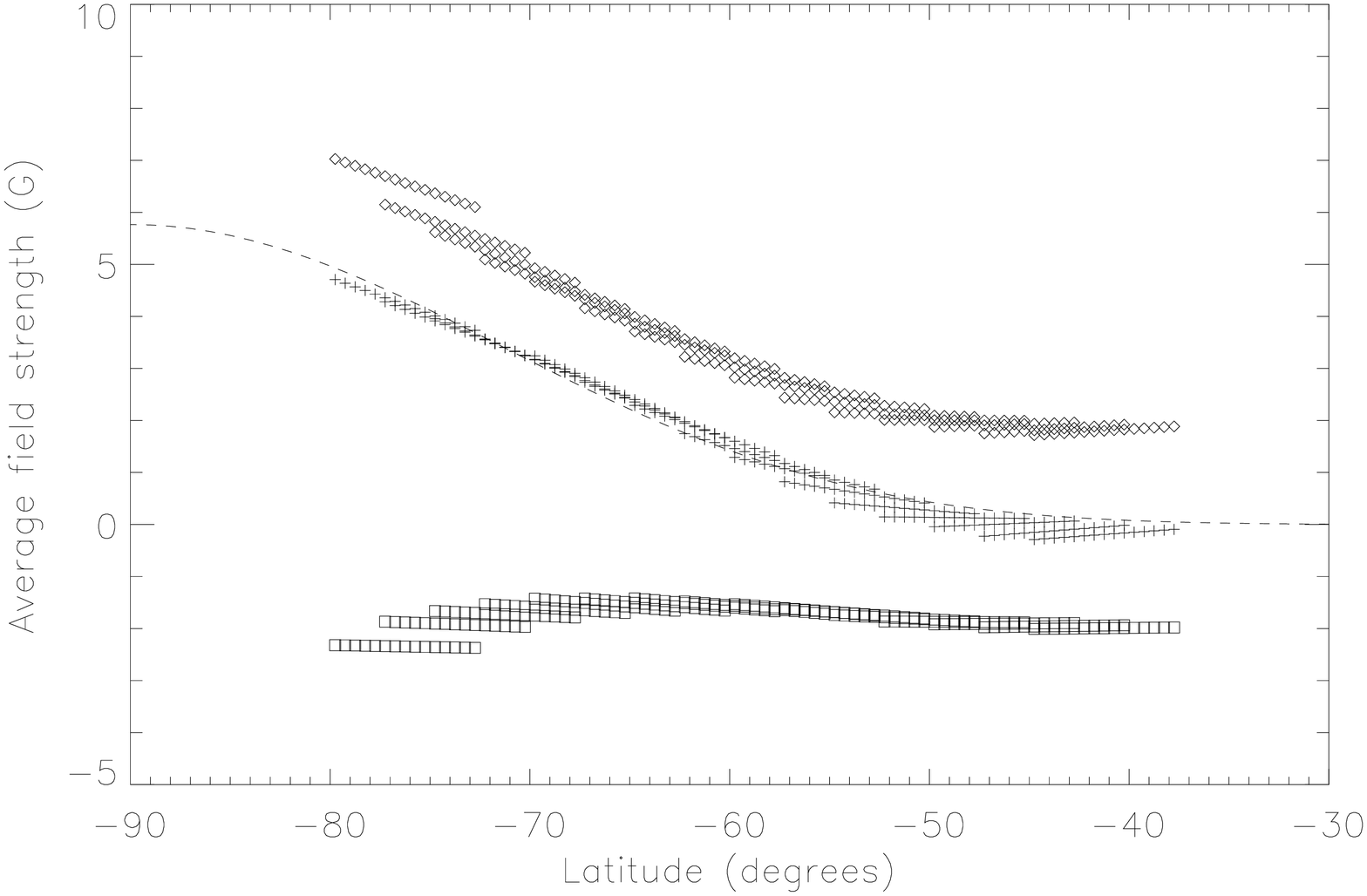}}
%\resizebox{0.89\hsize}{!}{\includegraphics*{keypoletilts.eps}}
\end{center}
\caption{Polar photospheric field distributions derived assuming that the field is approximately radial.  The net field data are fitted with a function of the form $B_{pole}\cos^n\theta$.  For the north pole, the best-fitting parameter values are $B_{pole} =-5.3$ and $n=8.8$, and for the south pole, $B_{pole} =5.8$ and $n=9.7$.}
\label{polefieldradial}
\end{figure*}

Having established that the photospheric polar fields are approximately radial, we next give our best estimate of the polar photospheric flux distribution by directly diagnosing the polar field assuming that it is radially directed.  We do this by averaging $B=B_r=B_{LOS}/\cos (\rho )$ within circles centered at selected positions on the disk such as those in Figure~\ref{fdcircles}.  Each straight line in each plot in Figure~\ref{polefieldradial} delineates a best linear fit to the set of month-by-month average measurements within a single circle over its range of latitudes.  Collected together in the plots, these straight lines form curves describing flux distributions from about $40^{\circ}$ and $80^{\circ}$.  The dashed lines show the best-fitting curves of the form $B_{pole}\cos^n\theta$ to the data, where $\theta$ is the co-latitude.  The best-fitting values of polar field strength $B_{pole}$ and power index $n$, given in the caption, are comparable to those found by previous authors using forward models for Cycle 21 (Svalgaard et al.~1978) and Cycle 22 (Wang \& Sheeley~1988).  The cosine(co-latitude) form may not be the best description of our Cycle-23 polar fields: the step up from background to polar field strength is more abrupt than in the cosine(co-latitude) model.  Furthermore, the high-latitude north-pole measurements show evidence of a plateau beyond $80^{\circ}$ while the south pole plot shows no such sign.  This may be related to the super-radial expansion of the south polar chromospheric field discussed above.  These are details that the forward modeling techniques used by past authors could not detect.

\section{Discussion}
\label{discussion}

Despite similar appearance of the chromopsheric and photospheric magnetograms in Figure~\ref{fulldiskimages}, there are major differences between the photospheric and chromospheric fields that become clear from the analysis of many measurements. We conclude that the magnetic fields in the solar photosphere and the chromosphere behave distinctly differently. Most of the photospheric field is within about $12^{\circ}$ of being radial, while the chromospheric field is more complex and has no strongly preferred direction, expanding in all directions to a significant degree. 

Besides the intrinsic scientific interest of solar surface magnetic field observations, their correct interpretation is of crucial importance to our ability to describe the magnetic structure of the solar atmosphere.  The recent advances in our understanding of photospheric field structure have brought into question the traditional picture of a photospheric field dominated by strong, nearly vertical flux associated with intergranular lanes.  It is now believed that these strong, intergranular fields are accompanied by an equally important, highly dynamic, nearly horizontal component associated with the internetwork.  For the purposes of synoptic construction and global solar atmospheric field modeling, the steady-state field orientation is desired, from which equilibrium field configurations are calculated.  In the results presented here, averaging the flux over many observations appears to reduce the effect of the horizontal fields, leaving the more steady, approximately radial flux to dominate the signal.  This same near-vertical flux is therefore likely to constitute the principal ingredient of synoptic magnetograms. 

The sensitivity of models to poorly observed or unobserved high-latitude fields motivated us to characterize the general properties of these fields as accurately as possible.  We did this not via forward models as in past studies but by inverting the stereoscopic projections of the field at a given latitude along different lines of sight at different times of year.  Significant longitudinal structure seemed to be absent from the sky images at polar latitudes.  The analysis treated latitude-by-latitude monthly averages and revealed poleward monotonic increases in intensity and unipolarity.  The photospheric polar field structure was approximately steady between 2003 and 2008 while the chromospheric field appears to be more time-dependent, becoming approximately steady only approaching activity minimum in 2008.  We found that the photospheric field is approximately radial in both polar caps between 2003 and 2008 and had field intensity increasing poleward, varying approximately as $B_{pole}\cos^n\theta$ with $B_{pole} =-5.3$ and $n=8.8$, and for the south pole, $B_{pole} =5.8$ and $n=9.7$, where $\theta$ is the co-latitude.  The chromospheric field tended to expand super-radially in the south during 2008.

In the context of synoptic magnetogram construction, these results support the prevailing policy in global solar atmospheric field modeling, where the photospheric LOS data are assumed to be projections of predominantly radial fields and are deprojected to form synoptic magnetograms for the radial field component.  The breakdown of the measurements beyond $80^{\circ}$ latitude corresponds to the 89th sine(latitude) pixel out of 90 in the standard $360\times 180$ longitude-sine(latitude) synoptic images.  The polar fields up to the 89th pixel are reasonably well determined to be approximately radial and to have a peaked distribution similar to a cosine(colatitude) pattern.  By our methods, the information needed to fill the 90th pixel is not available from the full-disk sky images.  Either observations from outside the ecliptic plane or models are needed to fill the gap.  As for the chromospheric fields, their lack of preferred direction suggests that any synoptic magnetogram formed from chromospheric images should be constructed in terms of the LOS field component, without assuming the field to be approximately radial.

%%%%%%%%%%%%%%%%%%%%%%%%%%%%%%%%%%%%%%%%%%%%%%%%%%%%%%%%

\acknowledgements
We thank the referee for a comprehensive report and Janet Luhmann for helpful comments.  The National Solar Observatory (NSO) is operated by the AURA, Inc., under cooperative agreement with the NSF. SOLIS data are produced cooperatively by NSF/NSO and NASA/LWS.  GP gratefully acknowledges support from NASA grant NNH05AA12I.  IP carried out this work through the National Solar Observatory Research Experiences for Undergraduate (REU) site program, which is cofunded by the Department of Defense in partnership with the National Science Foundation REU Program.

\end{document}